\documentclass[a4paper,11pt]{article}

\usepackage{epsfig,subfigure,cite}
\usepackage{amsmath}
\usepackage{macro,color}

\renewcommand{\Proofit}{{\it Proof\,:}\hspace{0.2cm}}

\begin{document}


\title{\LARGE
Distributed Randomized Algorithms for\\
the PageRank Computation\thanks{%
This paper is a preliminary version of the article that appeared in 
the IEEE Transactions on Automatic Control, 55: 1987-2002, 2010. 
This work was supported in part by the Ministry of Education, Culture,
Sports, Science and Technology, Japan, under KAKENHI Grant No.~21760323.}
}

\author{%
Hideaki Ishii\\[1mm]
Department of Computational Intelligence 
and Systems Science\\
Tokyo Institute of Technology\\
4259 Nagatsuta-cho, Midori-ku, Yokohama 226-8502, Japan\\
E-mail: ishii@dis.titech.ac.jp\\[4mm]
Roberto Tempo\\[1mm]
CNR-IEIIT, Politecnico di Torino\\
Corso Duca degli Abruzzi 24, 10129 Torino, Italy\\
E-mail: roberto.tempo@polito.it}

\addtolength{\abovedisplayskip}{-.4mm}
\addtolength{\belowdisplayskip}{-.4mm}
\addtolength{\jot}{-.2mm}
\addtolength{\textfloatsep}{-2mm}
\addtolength{\floatsep}{2mm}
\addtolength{\parsep}{12mm}
\setlength{\baselineskip}{3.4ex}


\maketitle

\begin{abstract}
In the search engine of Google, the PageRank algorithm plays 
a crucial role in ranking the search results.  The algorithm quantifies 
the importance of each web page based on the link structure of the web.
We first provide an overview of the original problem setup. 
Then, we propose several distributed randomized schemes
for the computation of the PageRank, where the pages
can locally update their values by communicating to those
connected by links.  
The main objective of the paper is to show that these schemes
asymptotically converge in the mean-square sense to the true PageRank
values. A detailed discussion on the close relations 
to the multi-agent consensus problems is also given. 
\end{abstract}


\section{Introduction}
\label{sec:intro}

In the last decade, search engines have become widely used
indispensable tools for searching the web.  For such engines, 
it is essential that the search results not only consist of web pages 
related to the query terms, but also rank the pages properly
so that the users quickly have access to the desired information. 
The PageRank algorithm at Google is one of the successful algorithms 
that quantify and rank the importance of each web page.
This algorithm was initially proposed in \cite{BriPag:98}, and
an overview can be found in, e.g., \cite{LanMey:06,BryLei:06}.

One of the main features of the PageRank algorithm is that it 
is based solely on the link structure inherent in the web.  
The underlying key idea is that links from important 
pages make a page more important.  More concretely, each page is considered to 
be voting the pages to which it is linked.  Then, in the ranking of a page,
the total number of votes as well as the importance of the voters 
are reflected.  This problem is mathematically formulated as finding 
the eigenvector corresponding to the largest eigenvalue of
a certain stochastic matrix associated with the web structure. 

For the PageRank computation, a critical aspect is the size of the web.
The web is said to be composed of over 8 billion pages, and its size
is still growing.  
Currently, the computation is performed centrally at Google, where the
data on the whole web structure is collected by crawlers 
automatically browsing the web.  
In practice, the class of algorithms that can be applied is limited.
In fact, the basic power method is employed, but it is reported that
this computation takes about a week \cite{LanMey:06}. 
This clearly necessitates more efficient computational methods.

In this regard, several approaches have recently been proposed.
In \cite{KamHavGol:04}, an adaptive computation method is developed,
which classifies web pages into groups based on the speed of 
convergence to the PageRank values and allocates computational
resources accordingly.
Another line of research is based on distributed approaches,
where the computation is performed on multiple servers communicating 
to each other. For example, Monte Carlo methods are used 
in \cite{AvrLitNem:07}, while the work in \cite{ZhuYeLi:05} 
utilizes the block structure of the web 
to apply techniques from the Markov chain literature.
In \cite{deJBra:07,KolGalSzy:06}, methods based on 
the so-called asynchronous iterations \cite{BerTsi:89} 
in numerical analysis are discussed.

In this paper, we follow the distributed approach and, in particular,
develop a randomized algorithm for the PageRank computation;
for recent advances on probabilistic approaches in systems and control, 
see \cite{TemCalDab_book}.
This algorithm is fully distributed and has three main features as follows:
First, in principle, each page can compute its own PageRank value locally
by communicating with the pages that are connected by direct links.
That is, each page exchanges its value with the pages that it is linked to and
those linked to it.  
Second, the pages make the decision to initiate this communication at
random times which are independent from page to page.
This means that, in its implementation,
there is neither a fixed order among the pages nor 
a centralized agent in the web that determines the pages to 
update their values.
Third, the computation required for each page is very mild.

The main result of the paper shows that the algorithm converges
to the true PageRank values in the mean-square sense.  
This is achieved by computing the time average at each page.
From a technical viewpoint, an important characteristic of the 
approach is that the stochasticity of the matrix in the original 
problem is preserved and exploited. 
We first propose a basic distributed update scheme for the
pages and then extend this into two directions to enhance 
its performance and flexibility for implementation. 
It is further noted that in \cite{IshTem_acc:09,IshTemBaiDab:09}, 
this approach has been generalized to incorporate failures 
in the communication as well as aggregation of the web structure. 
In \cite{IshTem_sice:09}, a related result on finding the 
variations in the PageRank values when the web data may 
contain errors is given. 

We emphasize that the approach proposed here is particularly motivated by 
the recent research on distributed consensus, agreement, and flocking problems 
in the systems and control community; see, e.g.,
\cite{JadLinMor:03,BerTsi:07,Wu:06,HatMes:05,BoyGhoPra:06,QuWanHul:08,TahJad:08,TemIsh:07,CarFagFoc:05,KasBasSri:07,YuHenFid:07,MarBroFra:04,RenBea:05,Moreau:05,LinFraMag:05}.
For additional details, we refer to 
\cite{AntBai:07,csm:07,BerTsi:89}.
Among such problems, our approach to PageRank computation is especially 
related to 
the consensus, where multiple agents exchange their values with neighboring 
agents so that they obtain consensus, i.e., all agents reach the same value.
The objective is clearly different from that of the PageRank problem, 
which is to find a specific eigenvector of a stochastic matrix
via the power method.
However, the recursion appearing in the consensus algorithm is exactly 
in the same form as the one for our distributed PageRank computation
except that the class of stochastic matrices is slightly different. 
These issues will be discussed further.

The organization of this paper is as follows: 
In Section~\ref{sec:pagerank}, we present an overview
of the PageRank problem. 
The distributed approach is introduced in Section~\ref{sec:dist1},
where we propose a basic scheme and prove its convergence. 
Its relation with multi-agent consensus
problems is discussed in Section~\ref{sec:consensus}.
We then develop two extensions of the basic distributed algorithm:
One in Section~\ref{sec:simul} is to improve the rate of convergence
by allowing multiple pages to simultaneously update and the other
in Section~\ref{sec:approx} to reduce the communication
load among the pages.
The proposed algorithm is compared with an approach 
known as asynchronous iteration
from the field of numerical analysis 
in Section~\ref{sec:asynch}.
Numerical examples are given in Section~\ref{sec:example} to show 
the effectiveness of the proposed schemes.
We conclude the paper in Section~\ref{sec:concl}.
Part of the material of this paper has appeared
in a preliminary form in \cite{IshTem_cdc:08}.

\medskip
\noindent
{\it Notation}:~~For vectors and matrices, inequalities 
are used to denote entry-wise inequalities:
For $X,Y\in\R^{n\times m}$, $X\leq Y$ implies
$x_{ij}\leq y_{ij}$ for $i=1,\ldots,n$ and $j=1,\ldots,m$;
in particular, we say that the matrix $X$ is nonnegative if $X\geq 0$ 
and positive if $X> 0$.
A probability vector is a nonnegative vector $v\in\R^n$ such that
$\sum_{i=1}^n v_i = 1$.
Unless otherwise specified,
by a stochastic matrix, we refer to a column-stochastic matrix,
i.e., a nonnegative matrix $X\in\R^{n\times n}$ with 
the property that $\sum_{i=1}^n x_{ij}=1$ for $j=1,\ldots,n$.
Let $\one\in\R^n$ be the vector with all entries equal to $1$ as
$\one:=[1\;\cdots\;1]^T$.
Similarly, $S\in\R^{n\times n}$ is the matrix with all entries being $1$.
For $x\in\R^n$, we denote by $\abs{x}$ the vector containing the
absolute values of the corresponding entries of $x$. 
The norm $\norm{\cdot}$ for vectors is the Euclidean norm.
The spectral radius of the matrix $X\in\R^{n\times n}$ is
denoted by $\rho(X)$. We use $I$ for the identity matrix.


\section{The PageRank problem}
\label{sec:pagerank}

In this section, we provide a brief introductory description of 
the PageRank problem; this material can be found in, e.g., 
\cite{LanMey:06,BryLei:06,BriPag:98}.

Consider a network of $n$ web pages indexed by integers from 1 to $n$.
This network is represented by the directed graph $\Gcal=(\Vcal,\Ecal)$.  
Here, $\Vcal:=\{1,2,\ldots,n\}$ is the set of vertices corresponding 
to the web page indices
while $\Ecal\subset\Vcal\times\Vcal$ is the set of edges representing
the links among the pages. 
The vertex $i$ is connected to the vertex $j$ by an edge, i.e.,
$(i,j)\in\Ecal$, if page $i$ has an outgoing link to page $j$,
or in other words, page $j$ has an incoming link from page $i$.
To avoid trivial situations, we assume $n\geq 2$.

The objective of the PageRank algorithm is to provide some
measure of importance to each web page.  
The PageRank value, or simply the value, of page $i\in\Vcal$
is a real number in $[0,1]$; we denote this by $x_i^*$. 
The values are ordered such that $x_i^*>x_j^*$ implies that page $i$ is 
more important than page $j$.  

The basic idea in ranking the pages in terms of the values is that
a page having links from important pages is also important.
This is realized by determining the value of one page as a sum
of the contributions from all pages that have links to it.  
In particular, the value $x_i^*$ of page $i$ is defined as
\[
  x_i^* = \sum_{j\in \Lcal_i} \frac{x_j^*}{n_j},
\]
where $\Lcal_i:=\{j:\; (j,i)\in\Ecal\}$, i.e., 
this is the index set of pages linked to page $i$,
and $n_j$ is the number of outgoing links of page $j$.
It is customary to normalize the total of all values 
so that $\sum_{i=1}^{n} x^*_i = 1$.

Let the values be in the vector form as $x^*\in[0,1]^n$.
Then, from what we described so far, the PageRank problem
can be restated as
\begin{equation}
   x^* = A x^*,~~x^*\in[0,1]^n,~~\sum_{i=1}^n x^*_i = 1,
\label{eqn:xA:pr}
\end{equation}
where the matrix $A=(a_{ij})\in\R^{n\times n}$, called the
link matrix, is given by
\begin{equation}
 a_{ij} 
  := \begin{cases}
      \frac{1}{n_j} & \text{if $j\in \Lcal_i$},\\
      0             & \text{otherwise}.
    \end{cases}
 \label{eqn:A}
\end{equation}
Note that the value vector $x^*$ is a nonnegative unit 
eigenvector corresponding to the eigenvalue 1 
of the nonnegative matrix $A$.
In general, for this eigenvector to exist and
to be unique, 
it is sufficient that the web as a graph is strongly connected
\cite{HorJoh:85}%
\footnote{A directed graph is said to be strongly connected
if for any two vertices $i,j\in\Vcal$, there is a sequence of
edges which connects $i$ to $j$.
In terms of the link matrix $A$, strong connectivity 
of the graph is equivalent to $A$ being irreducible.}.
However, the web is known not to be strongly connected.
Thus, the problem is slightly modified as follows%
\footnote{%
In fact, in the consensus literature, it is known that 
the eigenvalue 1 of a row-stochastic matrix is simple
if and only if the underlying graph has at least one globally reachable 
node; this means that there is a node from which
each node in the graph can be reached via a sequence of edges
(see, e.g., \cite{LinFraMag:05,RenBea:05,Moreau:05}). 
For our purpose, it is indeed possible to provide the 
column-stochastic counterpart of global reachable nodes, 
but the real web does not possess this property either.}.

First, note that in the real web, the so-called dangling nodes, 
which are pages having no links to others, are abundant.
Such pages can be found, e.g., in the form of PDF document files
having no outgoing links.
These pages introduce zero columns into the link matrix.
To simplify the discussion, we redefine the graph and thus the
matrix $A$ by bringing in artificial links for 
all dangling nodes (e.g., links back to the pages 
having links to a dangling node).
As a result, the link matrix $A$ becomes a stochastic matrix,
that is, $\sum_{i=1}^n a_{ij}=1$ for each $j$.
This implies that there exists at least one eigenvalue equal to 1. 

To emphasize the changes in the links that we have just made, 
we state the following as an assumption.

\begin{assumption}\rm
The link matrix $A$ given in \eqref{eqn:A}
is a stochastic matrix.
\end{assumption}

Next, to guarantee the uniqueness of the eigenvalue 1, 
a modified version of the values has been introduced 
in \cite{BriPag:98} as follows:  
Let $m$ be a parameter such that $m\in(0,1)$, 
and let the modified link matrix $M\in\R^{n\times n}$ be defined by
\begin{equation}
  M := (1-m)A + \frac{m}{n}S.
  \label{eqn:M}
\end{equation}
In the original algorithm in \cite{BriPag:98}, 
a typical value for $m$ is chosen as $m=0.15$; 
we use this value throughout this paper%
\footnote{In \cite{BriPag:98}, no specific reason is given
for this choice of $m$. As shown later in \eqref{eqn:lambda2}, however,
large $m$ has the effect of faster convergence in the computation
while it can also average out the PageRank values.}.
Notice that $M$ is a positive stochastic matrix. 
By the Perron theorem \cite{HorJoh:85},
this matrix is primitive%
\footnote{A nonnegative matrix $X\in\R^{n\times n}$ is said to 
be primitive if it is irreducible and has only one eigenvalue
of maximum modulus.};
in particular, the eigenvalue 1 is of multiplicity~1 and 
is the unique eigenvalue of maximum modulus
(i.e., with the maximum absolute value). Furthermore,
the corresponding eigenvector is positive. 
Hence, we redefine the value vector $x^*$ by using $M$ in place of $A$ 
in \eqref{eqn:xA:pr} as follows.

\begin{definition}\rm
The PageRank value vector $x^*$ is given by 
\begin{equation}
   x^* = M x^*,~~x^*\in[0,1]^n,~~\sum_{i=1}^n x^*_i = 1.
 \label{eqn:prvec}
\end{equation}
\end{definition}

As mentioned in the Introduction, 
due to the large dimension of the link matrix $M$, the computation
of the eigenvector corresponding to the eigenvalue 1 is difficult.  
The solution that has been employed in practice is based on the power method.
That is, the value vector $x^*$ is computed through the recursion 
\begin{align}
  x(k+1) 
    &= M x(k) 
     = (1-m)A x(k) + \frac{m}{n}\one,
\label{eqn:xM0} 
\end{align}
where $x(k)\in\R^n$ and 
the initial vector $x(0)\in\R^n$ is a probability vector. 
The second equality above follows from the fact 
$Sx(k)=\one$, $k\in\Z_+$.
For implementation, the form on the far right-hand side
is important,  
exhibiting that multiplication is
required using only the sparse matrix $A$, and not the dense 
matrix $M$.

Based on this method, we can asymptotically find the
value vector as shown below; see, e.g., \cite{HorJoh:85}.

\begin{lemma}\rm
In the update scheme \eqref{eqn:xM0},
for any initial state $x(0)$ that is a probability vector, 
it holds that $x(k)\rightarrow x^*$ as $k\rightarrow\infty$.
\end{lemma}

We now comment on the convergence rate of this scheme.
Denote by $\lambda_1(M)$ and $\lambda_2(M)$ 
the largest and the second largest eigenvalues of $M$ 
in magnitude. 
Then, for the power method applied to $M$, the 
asymptotic rate of convergence is exponential and
depends on the ratio $\abs{\lambda_2(M)/\lambda_1(M)}$.
Since $M$ is a positive stochastic matrix, 
we have $\lambda_1(M)=1$ and $\abs{\lambda_2(M)}<1$.
Furthermore, it is shown in \cite{LanMey:06} that 
the structure of the link matrix $M$ leads us to the bound 
\begin{equation}
  \abs{\lambda_2(M)} \leq 1-m.
  \label{eqn:lambda2}
\end{equation}

We next provide a simple example for illustration.

\begin{figure}
  \vspace*{2mm}
  \centering
  \resizebox{3cm}{!}{\input{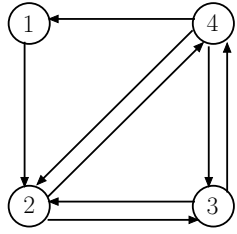}}\\[-2mm]
  \caption{A web with four pages}
  \label{fig:4nodes}
\end{figure}

\begin{example}\label{ex:1}\rm
Consider the web with four pages shown in Fig.~\ref{fig:4nodes}.
As a graph, this web is strongly connected, and there are no
dangling nodes. The link matrix $A$ and the modified link matrix $M$
can easily be constructed by \eqref{eqn:A} and \eqref{eqn:M}, 
respectively, as
\begin{align*}
  A &= \begin{bmatrix}
        0 & 0 & 0 & \frac{1}{3}\\
        1 & 0 & \frac{1}{2} & \frac{1}{3}\\
        0 & \frac{1}{2} & 0 & \frac{1}{3}\\
        0 & \frac{1}{2} & \frac{1}{2} & 0
      \end{bmatrix},~~
  M = \begin{bmatrix}
        0.0375 & 0.0375 & 0.0375 & 0.3208\\
        0.8875 & 0.0375 & 0.4625 & 0.3208\\
        0.0375 & 0.4625 & 0.0375 & 0.3208\\
        0.0375 & 0.4625 & 0.4625 & 0.0375
      \end{bmatrix},
\end{align*}
where we used the value $m=0.15$ from \cite{BriPag:98}.  
Then, the value vector $x^*$ can be computed as 
$x^* = \bigl[
          0.119 ~ 0.331 ~ 0.260 ~ 0.289
        \bigr]^T$.
Notice that page 2 has the largest value since it is linked 
from three pages while page 1, which has only one link to it, 
has the smallest value.
On the other hand, pages 3 and 4 have the same number of incoming links,
but page 4 has a larger value.  This is because page 4 has more
outgoing links, and thus it receives more contribution 
from page 3 than what it gives back.
\End
\end{example}

\section{A distributed randomized approach}
\label{sec:dist1}

In this section, we propose a distributed approach to compute 
the value vector $x^*$.

Consider the web 
from the previous section.
The basic protocol of the scheme is as follows:
At time $k$, page $i$ initiates its PageRank value
update (i) by sending the value of page $i$ to the pages that are linked 
and (ii) by requesting the values from the pages that are linked
to page $i$.
All pages involved in this process renew their values based on the latest
available information.  

To implement the scheme in a distributed manner, 
we assume that the pages taking the update action
are determined in a random manner.  
This is specified by the random process $\theta(k)\in\Vcal$, $k\in\Z_+$.
If at time $k$, $\theta(k)=i$, then page $i$ initiates an update
action by communicating and exchanging the values with
the pages connected by incoming and outgoing links.
Specifically, $\theta(k)$ is assumed to be i.i.d.,
and its probability distribution is given by
\begin{equation}
  \Prob\{\theta(k)=i\} = \frac{1}{n},~~\forall k\in\Z_+.
\label{eqn:theta}
\end{equation}
This means that each page takes the update action with
equal probability.  In principle, 
this scheme may be implemented without requiring 
a centralized decision maker or any fixed order among the pages.

In particular, consider the distributed update scheme in the following form:
\begin{equation}
  x(k+1) 
    = (1-\hat{m}) A_{\theta(k)} x(k) 
        + \frac{\hat{m}}{n}\one,
  \label{eqn:xMi1}
\end{equation}
where $x(k)\in\R^n$ is the state whose
initial state $x(0)$ is a probability vector, 
$\theta(k)\in\{1,\ldots,n\}$ is the mode of the system, and
$A_i$, $i=1,\ldots,n$, are called the distributed link matrices
and are to be determined;
$\hat{m}\in(0,1)$ is a parameter replacing $m$
in the centralized scheme \eqref{eqn:xM0}. 

The objective here is to design this distributed update scheme 
by finding the appropriate link matrices $A_i$ and the parameter $\hat{m}$
so that the PageRank values are computed
through the time average of the state $x$.
Let $y(k)$ be the average of the sample path $x(0),\ldots,x(k)$ as
\begin{equation}
  y(k) := \frac{1}{k+1}\sum_{\ell=0}^{k} x(\ell).
\label{eqn:yk}
\end{equation}
We say that, for the distributed update scheme, 
the PageRank value $x^*$ is obtained through the time average $y$ if,
for each initial state $x(0)$ that is a probability vector,
$y(k)$ converges to $x^*$ in the mean-square sense as follows:
\begin{equation}
 E\left[
   \bigl\|
      y(k) - x^*
   \bigr\|^2
  \right] \rightarrow 0,~~~k\rightarrow\infty.
 \label{eqn:thm:erg}
\end{equation}
This type of convergence 
is called ergodicity for stochastic processes \cite{PapPil:02}.

In what follows, we develop the distributed update scheme of \eqref{eqn:xMi1}. 
The main result is presented as Theorem~\ref{thm:erg}
showing the convergence of the scheme. 
The design consists of two steps, one for the link matrices $A_i$
and then the parameter $\hat{m}$. 
In later sections, this approach will be extended to improve the
convergence rate and the necessary communication load.

\subsection{Distributed link matrices and their average}

The first step in the development is 
to introduce the distributed link matrices.
For each $i$, the matrix $A_i\in\R^{n\times n}$ 
is obtained as follows: 
(i) The $i$th row and column coincide with those of $A$; 
(ii) the remaining diagonal entries are equal to $1-a_{i\ell}$,
$\ell=1,\ldots,n$, $\ell\neq i$; and
(iii) all the remaining entries are zero.
More formally, we have
\begin{align}
 (A_i)_{j\ell}
  &:= \begin{cases}
         a_{j\ell}   & \text{if $j=i$ or $\ell=i$},\\
         1-a_{i\ell} & \text{if $j=\ell\neq i$},\\
         0         & \text{otherwise},
     \end{cases} 
 ~~~~~i=1,\ldots,n.
 \label{eqn:Ai}
\end{align}
It follows that these matrices are stochastic
because the original link matrix $A$ possesses this property.
As we shall see later, this property indeed is
critical for the convergence of the scheme. 

\begin{example}\label{ex:3}\rm
We continue with the 4-page web in Example~\ref{ex:1}.
The link matrices $A_i$ 
are given by
\begin{align*}
 A_1 
  &= \begin{bmatrix}
       0           & 0 & 0 & \frac{1}{3}\\
       1           & 1 & 0 & 0\\
       0           & 0 & 1 & 0\\
       0           & 0 & 0 & \frac{2}{3}
     \end{bmatrix},~~
 A_2 
   = \begin{bmatrix}
       0           & 0           &           0 & 0\\
       1           & 0           & \frac{1}{2} & \frac{1}{3}\\
       0           & \frac{1}{2} & \frac{1}{2} & 0\\
       0           & \frac{1}{2} &           0 & \frac{2}{3}
     \end{bmatrix},~~
 A_3
  = \begin{bmatrix}
       1           & 0 & 0 & 0\\
       0           & \frac{1}{2} & \frac{1}{2} & 0\\
       0           & \frac{1}{2} & 0           & \frac{1}{3}\\
       0           &           0 & \frac{1}{2} & \frac{2}{3}
     \end{bmatrix},~~
 A_4
   = \begin{bmatrix}
       1           & 0           & 0 & \frac{1}{3}\\
       0           & \frac{1}{2} & 0 & \frac{1}{3}\\
       0           & 0           & \frac{1}{2} & \frac{1}{3}\\
       0           & \frac{1}{2} & \frac{1}{2} & 0
     \end{bmatrix}.
\tag*{\text{\End}}
\end{align*}

\end{example}


\medskip
To clarify the properties of the link matrices $A_i$ 
just introduced, we consider the simpler update scheme 
\begin{equation}
  x(k+1) = A_{\theta(k)} x(k),
\label{eqn:xAi}
\end{equation}
where $x(k)$ is the state with
$x(0)$ being a probability vector,
and the mode $\theta(k)$ is specified in \eqref{eqn:theta}.  
In particular, we focus on its average dynamics. 
The mean $\overline{x}(k):=E[x(k)]$ of the state $x(k)$ 
follows the recursion
$\overline{x}(k+1)=\overline{A} \overline{x}(k)$, where
$\overline{A}:= E[A_{\theta(k)}]$ is the average matrix
and $\E[\,\cdot\,]$ is the
expectation with respect to the random process $\theta(k)$.
Hence, we now inspect this matrix $\overline{A}$.
Due to the probability distribution of $\theta(k)$ 
in \eqref{eqn:theta}, we have
\begin{equation}
  \overline{A}
   = \frac{1}{n} \sum_{i=1}^n A_i. 
\label{eqn:Abar0}
\end{equation}
It is obvious that $\overline{A}$ is a stochastic matrix 
since all $A_i$ are stochastic.

The following lemma shows some properties of this matrix $\overline{A}$
that will be useful later. 

\begin{lemma}\label{lem:Abar}\rm
For the average matrix $\overline{A}$ given in \eqref{eqn:Abar0},
we have the following:
\begin{enumerate}
\item[(i)]  $\overline{A} = \frac{2}{n} A + \frac{n-2}{n} I$.
\item[(ii)] There exists a vector $z_0\in\R_+^n$ which is
an eigenvector corresponding to the eigenvalue 1 for both 
matrices $A$ and $\overline{A}$.
\end{enumerate}
\end{lemma}

\Proofit
(i)~~By definition of $\overline{A}$, we have
\begin{align*}
 (\overline{A})_{j\ell} 
  &= \frac{1}{n}
        \sum_{i=1}^{n} (A_i)_{j\ell}
  = \begin{cases}
       \frac{1}{n}
         \bigl[
          a_{jj} 
          + \sum_{i=1,\ i\neq j}^n
                    (1- a_{ij})
         \bigr] \\
         \hspace*{1.5cm} \text{if $j=\ell$},\\
       \frac{2}{n} a_{j\ell}~~~~\text{if $j\neq \ell$}.\\
     \end{cases}
\end{align*}
By definition of $A$, 
we have $a_{jj}=0$ and $\sum_{i=1,\ i\neq j}^n a_{ij} = 1$. 
Thus, the expression for $\overline{A}$ follows.

(ii)~~From (i), we have $\overline{A}-I = 2/n(A-I)$.  
This implies that 
every eigenvector $z_0$ of the link matrix $A$ associated with
the eigenvalue 1 
is also an eigenvector of the average matrix $\overline{A}$
for the same eigenvalue.
\EndProof

\smallskip
The lemma above provides some justification for the proposed distributed 
approach. That is, even though the matrices $A$ and $\overline{A}$ have 
different structures, they share an eigenvector for the eigenvalue 1,
which corresponds to the PageRank vector.

\subsection{Mean-square convergence of the distributed update scheme}

As in the case with the original link matrix $A$, 
for the average matrix $\overline{A}$, the eigenvector corresponding 
to the eigenvalue 1 may not be unique.  
We follow an argument similar to that in Section~\ref{sec:pagerank} and 
introduce the modified versions of the distributed link matrices. 

Since the link matrices $A_i$ are stochastic, we can rewrite
the distributed update scheme in \eqref{eqn:xMi1} as
\begin{equation}
  x(k+1) = M_{\theta(k)} x(k),
  \label{eqn:xM} 
\end{equation}
where the modified distributed link matrices are given by
\begin{equation}
  M_i 
   := (1-\hat{m}) A_i + \frac{\hat{m}}{n} S,~~~i=1,\ldots,n.
  \label{eqn:Mi}
\end{equation}
This expression is derived by the relation $Sx(k)=\one$
because $x(k)$ in \eqref{eqn:xMi1} 
is a probability vector for each $k$. 
Note that $M_i$ are positive stochastic matrices. 

Similarly to the argument on the link matrices $A_i$ above,
the problem at this second step is as follows: 
We shall find the modified link matrices $M_i$ by choosing the 
parameter $\hat{m}$ 
such that their average and the link matrix $M$ from \eqref{eqn:M} 
share an eigenvector for the eigenvalue 1. 
Since such an eigenvector is unique for $M$, it is necessarily 
equal to the value vector $x^*$ (see \eqref{eqn:prvec}).

Let $\overline{x}(k):=\E[x(k)]$ be
the mean of the state $x(k)$ of the system \eqref{eqn:xM}.
Its dynamics is expressed as
\begin{equation}
  \overline{x}(k+1) = \overline{M} \overline{x}(k),
\label{eqn:xMbar}
\end{equation}
where $\overline{x}(0)=x(0)$ and 
the average matrix $\overline{M}$ is given by
$\overline{M} := \E[M_{\theta(k)}]$.

A simple way of defining $M_i$ would be to let $\hat{m}=m$
as in the case with $M$;
however, it can be shown that there is no clear relation between
the original matrix $M$ and the average matrix $\overline{M}$ such as 
that between $A$ and $\overline{A}$ as we have seen in
Lemma~\ref{lem:Abar}.
Instead, we take the parameter $\hat{m}$ as 
\begin{equation}
   \hat{m} = \frac{2m}{n - m(n-2)}.
 \label{eqn:mhat}
\end{equation}
For the value $m=0.15$ used in this paper, 
we have $\hat{m} = 0.3/(0.85n+0.3)$. 
For this choice of $\hat{m}$, the next result holds.  

\begin{lemma}\label{lem:Mbar}\rm
For the parameter $\hat{m}$ given in \eqref{eqn:mhat}, 
we have the following:
\begin{enumerate}
\item[(i)] $\hat{m}\in(0,1)$ and $\hat{m}<m$.
\item[(ii)] 
$\overline{M} = \frac{\hat{m}}{m} M + \left(1 - \frac{\hat{m}}{m}\right)I$.
\item[(iii)]
For the average matrix $\overline{M}$, 
the eigenvalue 1 is simple and is the unique eigenvalue of maximum modulus.
The value vector $x^*$ is 
the corresponding eigenvector.
\end{enumerate}
\end{lemma}

\vspace*{2mm}
\Proofit
(i)~~By the assumptions $m\in(0,1)$ and $n\geq 2$, 
$\hat{m}$ in \eqref{eqn:mhat} is positive.  Also, we have
$1-\hat{m} = n(1-m)/[n - m(n-2)]$.
Hence, $1-\hat{m}> 1-m$, that is, $\hat{m}< m< 1$.

\noindent
(ii)~~This can be shown by direct calculation as follows:
\begin{align*}
 \overline{M}
   &= (1-\hat{m}) \overline{A} + \frac{\hat{m}}{n} S~~~
             \text{(by the definition of $\overline{M}$, 
                        \eqref{eqn:Mi}, 
                          and then \eqref{eqn:Abar0})}\\
   &= (1-\hat{m}) 
       \biggl[
          \frac{2}{n} A + \frac{n-2}{n} I 
       \biggr]
        + \frac{\hat{m}}{n} S~~~
           \text{(by Lemma~\ref{lem:Abar}\;(i))}\\
   &= \frac{\hat{m}}{m} M 
        + \biggl(
             1 - \frac{\hat{m}}{m}
          \biggr)I~~~
           \text{(by $\hat{m}$ in \eqref{eqn:mhat} and the 
                 definition of $M$ in \eqref{eqn:M})}.
\end{align*}

\noindent
(iii)~~From (ii), we have $\overline{M}-I = \hat{m}/m (M-I)$.
Hence, $\overline{M}$ and $M$ share an eigenvector
for the eigenvalue 1.  However, both $\overline{M}$ and $M$ 
are positive stochastic matrices.
Therefore, by the Perron theorem \cite{HorJoh:85},
for these matrices,
the eigenvalue 1 is of multiplicity 1 and is the unique one having the maximum
magnitude. Moreover, by \eqref{eqn:prvec}, the corresponding eigenvector
coincides with $x^*$.
\EndProof

\smallskip
From (iii) in the lemma above, it follows that the value vector $x^*$ 
can be obtained by the power method, i.e., by the average system 
in \eqref{eqn:xMbar} as
$\overline{x}(k) \rightarrow x^*$, $k\rightarrow\infty$.
Hence, in an average sense, the distributed update scheme 
asymptotically provides the correct values.  
It is interesting to observe that 
this can be achieved though the original link matrix $A$
does not explicitly appear in the scheme.
In fact, an eigenvector of the matrix $M$ is 
computed through randomly switching among the distributed link 
matrices $M_i$.

However, this property turns out not to be sufficient to guarantee 
convergence of $x(k)$ to the true value $x^*$. 
From Lemma~\ref{lem:tauM}\;(ii) in the Appendix, 
we can show that for any sequence $\{\theta(k)\}$,
there exists a sequence of probability vectors $\{v(k)\}$
such that, for any $x(0)$, 
it holds that $x(k)-v(k) \one^T x(0)=x(k)-v(k)\rightarrow 0$ 
as $k\rightarrow\infty$.  The vector $v$ and hence the state $x$ 
in general do not converge.
This can be seen in Example~\ref{ex:3} when, e.g., page~1 initiates
an update ($\theta(k)=1$); the update for page~4 
is given by $x_4(k+1) = 2(1-m)/3 x_4(k) + m/4$, showing that
$x_4$ cannot stay at its equilibrium value. 
Therefore, in the distributed approach, 
we resort to computing the time average $y(k)$ of the states.

The following theorem is the main result of this section.
It shows that the time average indeed converges to 
the value vector in the mean-square sense. 

\begin{theorem}\label{thm:erg}\rm
In the distributed update scheme in \eqref{eqn:xM},
the PageRank value $x^*$ is obtained through the 
time average $y$ in \eqref{eqn:yk} as 
$E\bigl[
   \bigl\|
      y(k) - x^*
   \bigr\|^2
  \bigr] \rightarrow 0$, $k\rightarrow\infty$.
\end{theorem}

\smallskip
The theorem highlights an ergodic property in 
the proposed update scheme.
It can be shown by general Markov process
results in, e.g., \cite{Cogburn:86}.
For completeness, however, 
a proof more specific to the current setup
is provided in Appendix~\ref{sec:app:B};
it employs tools for stochastic matrices and 
moreover is useful for an extension given in Section~\ref{sec:approx}.
Regarding the convergence of this algorithm, 
we see from \eqref{eqn:thm:erg:conv} in the proof
that it is of order $1/k$ and moreover
depends on the size of $n$ linearly through the parameter 
$\hat{m}$ in \eqref{eqn:mhat}. 

Several remarks are in order. 
In practice, each page needs to communicate with the 
pages that are directly connected by incoming or outgoing links.
We emphasize that the recursion 
to be
used is \eqref{eqn:xMi1} and not the equivalent expression of \eqref{eqn:xM};
in the latter case, the link matrices $M_i$ are positive,
which can imply that the values of all pages
are required for an update of a page. 
Nevertheless, as can be seen in \eqref{eqn:xMi1}, the link
matrices $A_i$ are sparse. Thus, at time $k$, communication 
is required only among the pages corresponding to the nonzero entries
in $A_{\theta(k)}$. 
Each page then performs weighted addition of its own value, the values 
just received, and the extra term $\hat{m}/n$.
Consequently, the amount of computation required 
at each page is limited at any time.

Implementation issues such as how web pages can exactly 
make local computations are outside the scope of this paper. 
However, it is clear that certain regulations 
may be necessary so that page owners cooperate with the 
search engine and the PageRank values computed by them 
can be trusted\footnote{%
In the consensus literature, problems involving cheating
have been studied. 
An example is the Byzantine agreement problem,
where among the agents there are malicious ones who send
confusing information so that other agents cannot achieve consensus
(see, e.g., \cite{TemIsh:07}).}.
Another issue concerning reliability of the ranking
is that of link spam, 
i.e., links added to enhance the PageRank of 
certain pages on purpose; a method to detect such 
spamming is studied in, e.g., \cite{AndBorHop:07,LanMey:06}.

\section{Relations to consensus problems}
\label{sec:consensus}

In this section, we discuss the relation between 
the two problems of PageRank and consensus.
First, we describe a stochastic version of the consensus problem.
Such problems have been studied
in, e.g., \cite{BoyGhoPra:06,Wu:06,HatMes:05,TahJad:08}; 
see also \cite{TemIsh:07}.

Consider a set $\Vcal=\{1,2,\ldots,n\}$ of agents having
scalar values.  
The network of agents is represented by the directed
graph $\Gcal=(\Vcal,\Ecal)$.
The vertex $i$ is connected to the vertex $j$ by an edge
$(i,j)\in\Ecal$ if agent $i$ can communicate its value to agent $j$.
Assume that the graph is strongly connected%
\footnote{As discussed in Section~\ref{sec:pagerank}, 
this assumption can be replaced with a weaker one that 
a globally reachable node exists.}. 

The objective is that all agents reach a common value
by communicating to each other, where
the pattern in the communication 
is randomly determined at each time.
Let $x_i(k)$ be the value of agent $i$ held at time $k$,
and let $x(k):=[x_1(k)\cdots x_n(k)]^T\in\R^n$. 
The values are updated via the recursion
\begin{equation}
  x(k+1) = A_{\theta(k)} x(k),
\label{eqn:consensus:x}
\end{equation}
where $\theta(k)\in\{1,\ldots,d\}$ is the mode 
specifying the communication pattern among the agents 
and $d$ is the number of such patterns. 
The communication patterns are given as follows:
Each $i\in\{1,\ldots,d\}$ corresponds to the subset
$\Ecal_i\subset\Ecal$ of the edge set. 
Then, the matrix $A_i$ has $(A_i)_{j\ell}>0$ if and only if
$(\ell,j)\in\Ecal_i$.
We assume that 
(i)~$(j,j)\in\Ecal_i$ for all $j$, (ii)~$\bigcup_{i=1}^d\Ecal_i=\Ecal$,
and (iii)~the matrix $A_{i}$ is a row-stochastic matrix.
The communication pattern is random, and in particular, 
$\theta(k)$ is an i.i.d.\ random process. Its probability 
distribution is given by 
$\Prob\{\theta(k)=i\} = 1/d$ for 
$k\in\Z_+$.

We say that consensus is obtained if 
for any initial vector $x(0)\in\R^n$, it holds that
\begin{equation}
  \abs{x_i(k)-x_j(k)}\rightarrow 0,~~k\rightarrow\infty
\label{eqn:consensus}
\end{equation}
with probability one for all $i,j\in\Vcal$.

A well-known approach is to update the value of each agent
by taking the average of the values received at that time.
In this case, the matrix $A_i$ is constructed as
\[
 (A_{i})_{j\ell}
  := \begin{cases}
      \frac{1}{n_{ij}} & \text{if $(\ell,j)\in\Ecal_{i}$},\\
      0                     & \text{otherwise},
    \end{cases}
\]
where 
$n_{ij}$ is the number of agents $\ell$ with $(\ell,j)\in\Ecal_{i}$,
i.e., those that transmit their values to agent $j$.

\begin{example}\label{ex:consensus}\rm
Consider the graph in Example~\ref{ex:1} with four agents.
We introduce four communication patterns arising from
the protocol in the distributed PageRank algorithm: 
The edge subset $\Ecal_i$ contains all $(i,j)$ and $(j,i)$
in the original edge set $\Ecal$ including
$(i,i)$ that corresponds to a self-loop for $i,j=1,2,3,4$.
The matrices $A_i$ can be written as
\begin{align*}
 A_1 
  &= \begin{bmatrix}
       \frac{1}{2} & 0 & 0 & \frac{1}{2}\\
       \frac{1}{2} & \frac{1}{2} & 0 & 0\\
       0           & 0 & 1 & 0\\
       0           & 0 & 0 & 1
     \end{bmatrix},~~
 A_2 
   = \begin{bmatrix}
       1           & 0           &           0 & 0\\
       \frac{1}{4} & \frac{1}{4} & \frac{1}{4} & \frac{1}{4}\\
       0           & \frac{1}{2} & \frac{1}{2} & 0\\
       0           & \frac{1}{2} &           0 & \frac{1}{2}
     \end{bmatrix},~~
 A_3
  = \begin{bmatrix}
       1           & 0 & 0 & 0\\
       0           & \frac{1}{2} & \frac{1}{2} & 0\\
       0           & \frac{1}{3} & \frac{1}{3} & \frac{1}{3}\\
       0           &           0 & \frac{1}{2} & \frac{1}{2}
     \end{bmatrix},~~
 A_4
   = \begin{bmatrix}
       \frac{1}{2} & 0           & 0 & \frac{1}{2}\\
       0           & \frac{1}{2} & 0 & \frac{1}{2}\\
       0           & 0           & \frac{1}{2} & \frac{1}{2}\\
       0           & \frac{1}{3} & \frac{1}{3} & \frac{1}{3}
     \end{bmatrix}.
\tag*{\text{\End}}
\end{align*}
\end{example}

We now present the convergence result for consensus.

\begin{lemma}\rm
Assume that the graph is strongly connected. Then under
the scheme of \eqref{eqn:consensus:x}, where the communication
pattern is chosen randomly, 
consensus is obtained in the sense of  \eqref{eqn:consensus}.
\end{lemma}

{\it Outline of the proof:}~~%
Let $\overline{A}:=E[A_{\theta(k)}]$ be the average matrix.
This matrix is stochastic and irreducible.
This is because the original graph is strongly 
connected, and hence under the probability distribution of $\theta(k)$,
we have $(\overline{A})_{j\ell}>0$ for each $(\ell,j)\in\Ecal$. 
Furthermore, by definition, the diagonal entries are positive, 
and thus $\overline{A}$ is a primitive matrix, implying that it has
the unique eigenvalue 1 of maximum modulus \cite{HorJoh:85}. 
Thus, by \cite{TahJad:08}, it follows that
consensus is obtained.
\EndProof

\smallskip
In comparison with the distributed PageRank problem, 
the consensus problem has the features below:
\begin{enumerate}
\item[(i)] The graph is assumed to be strongly connected. 
\item[(ii)] The goal is that all values $x_i(k)$ become equal,
and moreover there is no restriction on its size.
\item[(iii)] 
Convergence with probability one can be attained for the 
values $x_i(k)$ directly; there is no need to consider 
their time average (as in $y_i(k)$ in \eqref{eqn:yk}).
\item[(iv)] The matrices $A_i$ are row stochastic 
and the diagonal entries are all positive. 
In contrast, in our distributed PageRank computation scheme, 
the link matrices are column stochastic. 
However, the coefficient of ergodicity, which is the tool
employed for proving Theorem~\ref{thm:erg},
is useful also for this problem; see, e.g., \cite{TahJad:08}.
\end{enumerate}

It is clear that many similarities exist between the algorithms 
for consensus and PageRank.  We emphasize that the distributed
PageRank approach in this paper has been particularly motivated
by the recent advances in the consensus literature. 
We highlight two points that provide 
helpful insights into the PageRank problem as follows:
\begin{enumerate}
\item[(1)] At the conceptual level, it is natural to view
the web as a network of agents that can make its own computation
as well as communication with their neighboring agents.
\item[(2)] At the technical level, 
it is important to impose stochasticity on 
the distributed link matrices.  
For the distributed PageRank computation, very few
works exploit this viewpoint.
\end{enumerate}


\section{Extensions to simultaneous updates}
\label{sec:simul}

So far, we have discussed the update scheme
where only one page initiates an update at each time
instant. In the web with billions of pages, however,
this approach may not be practical.
In this section, we extend the distributed algorithm by
allowing multiple pages to simultaneously initiate updates.

Consider the web with $n$ pages from 
Sections~\ref{sec:pagerank} and \ref{sec:dist1}.
As before, at each time $k$, the page $i$ initiates its PageRank value
update (i) by sending its value to the pages that it is linked to 
and (ii) by requesting the pages that link to it for their values. 
All pages involved here update their values based on the new
information.
The difference from the simpler scheme before is that
there may be pages that are requested for their values 
by multiple pages at the same time.
The current scheme handles such situations.

These updates can take place in a fully distributed and 
randomized manner.  The decision to make an update is 
a random variable. In particular, this is determined under
a given probability $\alpha\in(0,1]$ at each time $k$, and
hence, the decision can be made locally at each page. 
Note that the probability $\alpha$ is a global parameter
in that all pages share the same $\alpha$.

Formally, the proposed distributed update scheme is described as follows.
Let $\eta_i(k)\in\{0,1\}$, $i=1,\ldots,n$, $k\in\Z_+$, 
be Bernoulli processes given by
\[
  \eta_i(k) 
   = \begin{cases}
       1 & \text{if page $i$ initiates an update at time $k$},\\
       0 & \text{otherwise},
     \end{cases}
\]
where their probability distributions are specified as
\begin{equation}
  \alpha := \Prob\bigl\{
                  \eta_i(k)=1
                \bigr\}.
\label{eqn:alpha}
\end{equation}
The process $\eta_i(k)$ is generated at the corresponding page $i$. 


Similarly to the argument in Section~\ref{sec:dist1}, 
we start with the update law as in \eqref{eqn:xMi1}:
\begin{equation}
  x(k+1) = (1-\hat{m}) A_{\eta_1(k),\ldots,\eta_n(k)} x(k)
             + \frac{\hat{m}}{n} \one,
\label{eqn:xAeta}
\end{equation}
where $x(k)\in\R^n$, the initial state $x(0)$ 
is a probability vector,
$\hat{m}\in(0,1)$ is the parameter used instead of $m$
in the centralized case,
and $A_{\eta_1(k),\ldots,\eta_n(k)}$ are the 
distributed link matrices.  

The problem of distributed PageRank computation is formulated
as follows: Find the distributed link matrices 
$A_{p_1,\ldots,p_n}$ and the parameter $\hat{m}$ 
such that, 
for the corresponding distributed update scheme \eqref{eqn:xAeta}, 
the PageRank value $x^*$ is obtained through the time average. 
This problem is a generalization of that in Section~\ref{sec:dist1},
where 
only one page initiates an update at a time.
The current approach is called the distributed scheme
with simultaneous updates.  Its analysis 
is more involved as we shall see.

\subsection{Distributed link matrices and their average}

We introduce the distributed link matrices. 
Let the matrices $A_{p_1,\ldots,p_n}$ be given by 
\begin{align}
 &\bigl(
    A_{p_1,\ldots,p_n}
 \bigr)_{ij} 
 := \begin{cases}
      a_{ij} & \text{if $p_i = 1$ or $p_j = 1$},\\
      1 - \sum_{h:~p_h=1} a_{hj}
              & \text{if $p_i = 0$ and $i=j$},\\
      0       & \text{if $p_i = p_j=0$ and $i\neq j$}
    \end{cases} 
\label{eqn:Aeta}
\end{align}
for $p_r\in\{0,1\}$, $r\in\{1,\ldots,n\}$, and $i,j\in\{1,\ldots,n\}$.
Clearly, there are $2^n$ matrices. 
They have the property that (i)~if $p_i=1$, then
the $i$th column and the $i$th row are the same as those in 
the original link matrix $A$,
(ii)~if $p_i=0$, then the $i$th diagonal entry is chosen so that
the entries of the $i$th column add up to 1, and
(iii)~all other entries are 0.
Hence, these are constructed as stochastic matrices.  
Notice that the link matrix $A_{p_1,\ldots,p_n}$ 
coincides with $A_i$ in \eqref{eqn:Ai}
when $p_i=1$ and $p_j=0$ for all $j\neq i$.


We next analyze the average dynamics of the 
distributed update scheme in \eqref{eqn:xAeta}.
For simplicity, as in Section~\ref{sec:dist1},
we use the same notation
$\overline{A}$ for the average link matrix given by
\begin{equation}
  \overline{A}:= E\bigl[A_{\eta_1(k),\ldots,\eta_n(k)}\bigr],
  \label{eqn:Abar}
\end{equation}
where $E[\,\cdot\,]$ is the expectation with respect to 
$\eta_i(k)$, $i\in\Vcal$. 
This matrix $\overline{A}$ is stochastic.

The following result shows that the average link 
matrix $\overline{A}$ has a clear relation to 
the original link matrix $A$. In particular, it
implies that the two matrices share the eigenvector for eigenvalue 1. 

\begin{proposition}\label{prop:simul:Aave}\rm
For the average link matrix $\overline{A}$ given in \eqref{eqn:Abar},
we have the following:
\begin{enumerate}
\item[(i)] $\overline{A} = \bigl[1-(1-\alpha)^2\bigr] A + (1-\alpha)^2 I$.
\item[(ii)] There exists a vector $z_0\in\R_+^n$ which is
an eigenvector corresponding to the eigenvalue 1 for both 
$A$ and $\overline{A}$.
\end{enumerate}
\end{proposition}

\vspace*{1mm}
The proof of this proposition is preceded by a preliminary
result. Observe that $\overline{A}$ can be written as
\begin{equation}
  \overline{A}
   = \sum_{\ell=0}^{n} \alpha^{\ell} (1-\alpha)^{n-\ell} \hat{A}_{\ell},       
 \label{eqn:simul:Aave}
\end{equation}
where the matrices $\hat{A}_{\ell}$, $\ell=0,1,\ldots,n$, are given by
\begin{equation}
  \hat{A}_{\ell}
   := \sum_{\substack{p_r\in\{0,1\},~r=1,\ldots,n:\;\\
                        \sum_{r=1}^{n}p_r = \ell}}
          A_{p_1,\ldots,p_n}.
\label{eqn:Ahatl}
\end{equation}
The matrix $\hat{A}_{\ell}$ is the sum of matrices 
for the cases where $\ell$ pages simultaneously initiate updates.  

These matrices can be explicitly written in terms of the
original link matrix $A$.  
Here, we use the binomial coefficient 
given by $\comb{r}{k}:=r!/[(r-k)!\;k!]$.  
Note that $\comb{r}{0}=1$ for any $r\in\Z_+$.

\begin{lemma}\label{lem:Ahatl}\rm
The matrices $\hat{A}_{\ell}$, $\ell=0,1,\ldots,n$, in \eqref{eqn:Ahatl}
can be expressed as follows:
\begin{equation}
  \hat{A}_{\ell}
   = \begin{cases}
       A  & \text{if $\ell=n$},\\
       nA & \text{if $\ell=n-1$},\\
      \bigl(\comb{n}{\ell} - \comb{n-2}{\ell}\bigr) A + \comb{n-2}{\ell} I
          & \text{if $\ell=0,\ldots,n-2$}.
     \end{cases}
\label{eqn:lem:Ahatl}
\end{equation}
\end{lemma}

\Proofit
We consider four cases as follows.

(1)~$\ell=n$:~~This is the case when all pages initiate 
updates, and thus by definition 
$\hat{A}_n = A_{1,\ldots,1} = A$.

(2)~$\ell=n-1$:~~When all but one page initiate updates,
it is obvious from the definition that
$A_{0,1,\ldots,1} = A_{1,0,1,\ldots,1} = \cdots 
= A_{1,\ldots,1,0} = A$.
Since there are $n$ such cases, their sum is 
$\hat{A}_{n-1} = nA$.

(3)~$\ell=0$:~~In the case when none of the pages initiates an update,
by definition, the matrix $A_{0,\ldots,0}$ reduces to the identity 
matrix as
$\hat{A}_0=A_{0,\ldots,0}=I$. 
Noting that $\comb{n}{0}=\comb{n-2}{0}=1$, we have
\eqref{eqn:lem:Ahatl}.

(4)~$\ell=1,\ldots,n-2$:~~To prove the
expression of $\hat{A}_{\ell}$ for these cases, we must show for each 
entry that
\begin{align}
  \bigl(
    \hat{A}_{\ell}
  \bigr)_{ij}
   &= \begin{cases}
       \bigl(
          \comb{n}{\ell} - \comb{n-2}{\ell}
       \bigr) a_{ij} & \text{if $i\neq j$},\\
       \comb{n-2}{\ell}    & \text{if $i= j$},
     \end{cases}\notag\\
   &\hspace*{2cm}i,j\in\{1,\ldots,n\}.  
\label{eqn:Ahatl:4}
\end{align}
In the following, the proof
is divided into two steps for the cases of $i\neq j$ and $i=j$.

(i)~$i\neq j$:~~By the definition of $A_{p_1,\ldots,p_n}$
in \eqref{eqn:Aeta}, its $(i,j)$ entry reduces to
\[
 \bigl(
   A_{p_1,\ldots,p_n}
 \bigr)_{ij} 
  = \begin{cases}
      a_{ij} & \text{if $p_i = 1$ or $p_j = 1$},\\
      0       & \text{otherwise}.
    \end{cases}   
\]
Thus, from \eqref{eqn:Ahatl}
\begin{align}
  \bigl(
    \hat{A}_{\ell}
  \bigr)_{ij}
  &= \sum_{\substack{p_r\in\{0,1\},~r=1,\ldots,n:\\
                        \sum_{r=1}^{n}p_r = \ell}}
          \bigl(
             A_{p_1,\ldots,p_n}
          \bigr)_{ij} \notag\\
  &= \sum_{\substack{p_r\in\{0,1\},~r=1,\ldots,n:\\
                        \sum_{r=1}^{n}p_r = \ell~\text{and}\\
                      (p_i = 1~\text{or}~p_j = 1)}}
           a_{ij} \notag\\
  &= \bigl(
        2 \comb{n-1}{\ell-1} - \comb{n-2}{\ell-2}       
     \bigr) a_{ij}, 
\label{eqn:lem:Ahatl:4a}
\end{align}
where the last equality is established by counting
the number of cases where $p_i=1$ or $p_j=1$ holds
among all possible combinations of 
$p_1,\ldots,p_n\in\{0,1\}$ such that $\ell$ of them are equal to 1.
Using the formula for binomial coefficients
\begin{equation}
  \comb{r}{k}=\comb{r-1}{k} + \comb{r-1}{k-1},~~~r,k\in\Z_+,
\label{eqn:binomial}
\end{equation}
we can show that
$2 \comb{n-1}{\ell-1} - \comb{n-2}{\ell-2}
 = \comb{n}{\ell} - \comb{n-2}{\ell}$.
Hence, 
from \eqref{eqn:lem:Ahatl:4a},
we arrive at \eqref{eqn:Ahatl:4} 
for $i\neq j$.

(ii)~$i=j$:~~Since $a_{ii}=0$ in the link matrix $A$
in \eqref{eqn:A} and by \eqref{eqn:Aeta},
the $(i,i)$ entry of $A_{p_1,\ldots,p_n}$ 
is
\[
 \bigl(
    A_{p_1,\ldots,p_n}
 \bigr)_{ii} 
  = \begin{cases}
      1 - \sum_{h:~p_h=1} a_{hi}
              & \text{if $p_i = 0$},\\
      0 & \text{if $p_i = 1$}.
    \end{cases}   
\]
Hence, 
\begin{align*}
  \bigl(
    \hat{A}_{\ell} 
  \bigr)_{ii}
  &= \sum_{\substack{p_r\in\{0,1\},~r=1,\ldots,n:\\
                        \sum_{r=1}^{n}p_r = \ell}}
          \bigl(
             A_{p_1,\ldots,p_n}
          \bigr)_{ii} 
  = \sum_{\substack{p_r\in\{0,1\},~r=1,\ldots,n:\\
                        \sum_{r=1}^{n}p_r = \ell,~p_i = 0}}
       \biggl(
         1 - \sum_{h:\;p_h=1} a_{hi}
       \biggr)\\
  &= \comb{n-1}{\ell}
      - \sum_{\substack{h=1\\ h\neq i}}^{n}
          \sum_{\substack{p_r\in\{0,1\},~r=1,\ldots,n:~\\
                \sum_{r=1}^{n}p_r = \ell,\;p_i = 0,\;p_h=1}}          
                    a_{hi},
\end{align*}
where the first term is obtained by counting the number of possible
combinations of $p_1,\dots,p_n$ such that their sum equals
$\ell$ and $p_i=0$; the second term is a consequence
of switching the order of two summations.
By a combinatorial argument again, we have
$\bigl(
    \hat{A}_{\ell} 
  \bigr)_{ii}
  = \comb{n-1}{\ell}
      - \comb{n-2}{\ell-1} \sum_{h=1, h\neq i}^n a_{hi}$.
The original link matrix $A$ in \eqref{eqn:A} is stochastic with 
diagonal entries being 0.
This fact together with \eqref{eqn:binomial} 
yields
$\bigl(
    \hat{A}_{\ell} 
  \bigr)_{ii}
  = \comb{n-1}{\ell}
         - \comb{n-2}{\ell-1}
  = \comb{n-2}{\ell}$.
Therefore, \eqref{eqn:Ahatl:4}  is attained for the 
case $i=j$.
\mbox{}
\EndProof

\smallskip
{\it Proof of Proposition~\ref{prop:simul:Aave}:}~
(i)~~By \eqref{eqn:simul:Aave} and Lemma~\ref{lem:Ahatl}, 
the matrix $\overline{A}$ can be computed directly as
\begin{align}
  \overline{A}
   &= \sum_{\ell=0}^n \alpha^{\ell} (1-\alpha)^{n-\ell} \hat{A}_{\ell}\notag\\
   &= \sum_{\ell=0}^{n-2} \alpha^{\ell} (1-\alpha)^{n-\ell} 
       \bigl[        
           \bigl(
             \comb{n}{\ell} - \comb{n-2}{\ell}
           \bigr) A
             + \comb{n-2}{\ell} I  
       \bigr] 
        + \alpha^{n-1} (1-\alpha) n A
        + \alpha^{n} A \notag\\
   &= \sum_{\ell=0}^{n} \alpha^{\ell} (1-\alpha)^{n-\ell}
                                   \comb{n}{\ell}A 
       + \sum_{\ell=0}^{n-2} \alpha^{\ell} (1-\alpha)^{n-\ell} \comb{n-2}{\ell}
          \bigl( 
            I - A 
          \bigr). 
\label{eqn:prop:simul:1}  
\end{align}
In the first term above, we have by the binomial identity
$\sum_{\ell=0}^{n} \alpha^{\ell} (1-\alpha)^{n-\ell} \comb{n}{\ell}
  = [ \alpha + (1-\alpha)]^n 
  = 1$.
Similarly, for the second term, 
$\sum_{\ell=0}^{n-2} \alpha^{\ell} (1-\alpha)^{n-\ell} \comb{n-2}{\ell}
    = (1-\alpha)^2$.
Substituting these relations into \eqref{eqn:prop:simul:1},
we have $\overline{A} = A + (1-\alpha)^2 (I - A)$,
which is the desired expression of 
$\overline{A}$.

(ii)~~The equality in (i) 
implies that any eigenvector $z_0$ of the link matrix $A$ associated with
the eigenvalue 1 
is also an eigenvector of 
the average matrix $\overline{A}$ for this eigenvalue.
\EndProof

\subsection{Mean-square convergence of the distributed update scheme}


To guarantee that the distributed scheme yields
the PageRank value, we now examine the modified versions
of the distributed link matrices. 
We first express the distributed update scheme of 
\eqref{eqn:xAeta} in its equivalent form as
\begin{align}
  x(k+1)
   &= M_{\eta_1(k),\ldots,\eta_n(k)} x(k),
  \label{eqn:xMmod}
\end{align}
where the modified distribution link matrices are given by
\begin{align}
  M_{p_1,\ldots,p_n}
  &:= (1-\hat{m}) A_{p_1,\ldots,p_n} 
        + \frac{\hat{m}}{n}S,
    ~~~~p_1,\ldots,p_n\in\{0,1\}.
\label{eqn:Meta}
\end{align}
This form can be obtained by using the facts that
the link matrices $A_{p_1,\ldots,p_n}$
are stochastic and that $Sx(k)=\one$. 
Clearly, these matrices $M_{p_1,\ldots,p_n}$ are
positive and stochastic. 


The objective here is to find the modified link 
matrices $M_{p_1,\ldots,p_n}$, by 
selecting the parameter $\hat{m}$,
so that their average and the link matrix $M$ from \eqref{eqn:M} 
share an eigenvector corresponding to the eigenvalue 1. 
Since such an eigenvector is unique for $M$, it is necessarily 
equal to the value vector $x^*$.


As in the earlier case in Section~\ref{sec:dist1},
we take the parameter $\hat{m}$ to be different from 
the original $m$. In particular, let
\begin{align}
  \hat{m} 
   = \frac{m[1 - (1 - \alpha)^2]}{1 - m (1 - \alpha)^2}.
 \label{eqn:mhat2}
\end{align}
For the value $m=0.15$ used in this paper, 
we have $\hat{m} 
= 0.15[1 - (1 - \alpha)^2]/[1 - 0.15 (1 - \alpha)^2]$.
Then, let the average link matrix be 
$\overline{M}:= \E[M_{\eta_1(k),\ldots,\eta_n(k)}]$.
Here, the distributed link matrices
are positive stochastic matrices, 
which implies that the average matrix $\overline{M}$ 
enjoys the same property.

The next lemma is the key to establish the desired
relation between the distributed link matrices and
their average. 
It is stated without proof; it follows similarly to that 
for Lemma~\ref{lem:Mbar}.

\begin{lemma}\label{lem:Mbar2}\rm
The scalar $\hat{m}$ in \eqref{eqn:mhat2} 
and the link matrices $M_{p_1,\ldots,p_n}$
in \eqref{eqn:Meta} have the following properties:
\begin{enumerate}
\item[(i)] $\hat{m}\in(0,1)$ and $\hat{m}\leq m$.
\item[(ii)]
$\overline{M} = \frac{\hat{m}}{m} M + \left(1 - \frac{\hat{m}}{m}\right)I$.
\item[(iii)] 
For the average matrix $\overline{M}$, 
the eigenvalue 1 is simple and is the unique eigenvalue of maximum modulus.
The value vector $x^*$ is 
the corresponding eigenvector.
\end{enumerate}
\end{lemma}

\smallskip
We can show by (iii) in the lemma that, in an average sense, 
the distributed update scheme asymptotically obtains the correct 
values.  More precisely, we have
$E[x(k)]=\overline{M}^{k} x(0)\rightarrow x^*$ 
as $k\rightarrow\infty$.
Further, as discussed in Section~\ref{sec:pagerank},
the asymptotic rate of convergence is dominated by
the second largest eigenvalue $\lambda_2(\overline{M})$ 
in magnitude. By \eqref{eqn:lambda2} and
(ii) in the lemma, this eigenvalue can be bounded as
\begin{align*}
  \abs{\lambda_2(\overline{M})}
    &= \frac{\hat{m}}{m}\abs{\lambda_2(M)} 
       + 1 - \frac{\hat{m}}{m}
    \leq \frac{1 - m}{1 - m (1 - \alpha)^2}.
\end{align*}
It is clear that this is a monotonically decreasing 
function of $\alpha$ and $m$.  That is, higher probability in updates
and/or larger $m$ results in faster convergence in average.

We are now ready to state the main result of this section. 

\begin{theorem}\label{thm:erg2}\rm
Consider the distributed scheme with simultaneous updates 
in \eqref{eqn:xMmod}.  For any update probability $\alpha\in(0,1]$, 
the PageRank value $x^*$ is obtained through the time average $y$
in \eqref{eqn:yk} as
$E\bigl[
   \bigl\|
      y(k) - x^*
   \bigr\|^2
  \bigr] \rightarrow 0$, $k\rightarrow\infty$.
\end{theorem}

\vspace*{1mm}
The proof follows along similar lines as that in 
Theorem~\ref{thm:erg}.  More specifically, we can prove
either by the general Markov chain results of, e.g.,
\cite{Cogburn:86} or by Appendix~\ref{sec:app:B}
where we replace the expression of $\overline{M}$ there
with the one in Lemma~\ref{lem:Mbar2}. 
Hence, the convergence is of order
$1/k$, as in the algorithm of Section~\ref{sec:dist1};
it also depends on the update probability $\alpha$
but is independent of $n$.

We remark that this scheme is fully decentralized when $\alpha<1$.
It is parameterized by $\alpha$,
which determines the frequency in the updates,
communication load among the pages,
and the rate of convergence in the mean
as we have seen above.  
In practice, the recursion in \eqref{eqn:xMmod}
must be implemented in the equivalent form \eqref{eqn:xAeta}.
It is clear that communication is required only 
over the links corresponding to the nonzero entries 
in the link matrices there.
Each page then computes weighted additions of its own value, 
the values received from others, 
and the constant $\hat{m}/n$.
On the other hand, when $\alpha=1$, the scheme 
reduces to the original centralized one in Section~\ref{sec:pagerank}.
In this case, the distributed link matrix is $M_{1,\ldots,1}$
and coincides with the original $M$
because $\hat{m}=m$ and $A_{1,\ldots,1}=A$ from 
Lemma~\ref{lem:Ahatl}.
%

\section{Update termination in PageRank computation}
\label{sec:approx}

In this section, we further develop
the distributed algorithm for calculating the PageRank.  
We relax the objective and aim at obtaining approximate values 
of the PageRank. 
The key feature here is to allow the pages to terminate their updates 
at the point when the values have converged to a certain level. 
The benefit is that such values need to be transmitted only once 
to the linked pages; hence, the computation and 
communication load can be reduced. 

In a centralized setting, 
the idea of update termination for the PageRank computation 
has been introduced by \cite{KamHavGol:04}. 
We extend this idea
to the distributed update scheme of Section~\ref{sec:simul}.
First, we consider a simple case to attain a convergence result.
Then, we provide the details of the proposed algorithm.

\subsection{Convergence properties for the distributed scheme}

Consider the distributed update scheme with simultaneous updates
in \eqref{eqn:xAeta} for computing the
values $x(k)$ together with their time average $y(k)$.
Within this subsection, we fix the sample 
paths $\{\eta_j(k)\}_{k=0}^{k_0-1}$, $j=1,\ldots,n$, 
up to time $k_0-1$ of the processes specifying 
the updates in the pages. 
Suppose that some of the time averages $y_i(k_0)$ 
have, in an approximate sense, converged.  This is measured by
finding those that have varied only within sufficiently small
ranges for a certain number of time steps.
We introduce two parameters:
Let $\delta\in(0,1)$ be the relative error level, and
let $N_s$ be the number of steps.  
Using the history of its own time average $y_i$,
each page $i$ then determines at time $k_0$
whether the following condition holds:
\begin{equation}
 \abs{y_i(k_0) - y_i(k_0-\ell)} \leq \delta y_i(k_0),~~\ell=1,2,\ldots,N_s.
\label{eqn:yconv}
\end{equation}
If so, then (i)~the page $i$ will terminate its update and
fix its estimate at $y_i(k_0)$, and then (ii) this value $y_i(k_0)$
is transmitted to the pages connected to $i$ by direct links.
After this point, these values will be used at 
the pages performing further updates. 

The question of interest is whether 
the pages that continue with their updates after time $k_0$
will reach a good estimate of their true values.
In what follows, we show that the answer is positive
and the approximation level achievable 
in the estimate will be as good as that for the pages 
that have terminated their updates. 
Note that the analysis is based on the given 
sample paths $\{\eta_j(k)\}_{k=0}^{k_0-1}$, and hence
the state $x$ and its average $y$ up to time $k_0$ are fixed;
we study their stochastic behaviors after this time. 

Let $\Ccal(k_0)$ be the set of page indices that have reached 
good estimates at time $k_0$ as
\begin{align*}
 \Ccal(k_0) 
   &:= \bigl\{
       i\in\Vcal:~
         \abs{y_i(k_0) - y_i(k_0-\ell)} \leq \delta y_i(k_0),
           ~~\ell=1,2,\ldots,N_s
      \bigr\}.
\end{align*}
The cardinality of this set is denoted by $n_1(k_0)$. 
We assume $n_1(k_0)\geq 1$.
Also, let 
$\Ncal(k_0) := \Vcal \setminus \Ccal(k_0)$.

Based on these sets, 
we introduce a coordinate transformation for the state $x(k)$
and partition it as
\[
 x(k) 
   = \begin{bmatrix}
       x_{\Ccal}(k)\\ 
       x_{\Ncal}(k)
     \end{bmatrix},~~~k\geq k_0,
\]
where $x_{\Ccal}(k)\in\R^{n_1(k_0)}$ contains
the values of the pages in $\Ccal(k_0)$ and 
$x_{\Ncal}(k)\in\R^{n-n_1(k_0)}$ contains those 
of the pages in $\Ncal(k_0)$.
With slight abuse of notation, we write the transformed 
state by $x(k)$.
Also, we use the shorthand notation $A_{p}$ for 
$A_{p_1,\ldots,p_n}$, $p_i\in\{0,1\}$, $i\in\Vcal$.
Then, the distributed link matrices $A_{p}$ in \eqref{eqn:Aeta}
and the average link matrix $\overline{A}$ in \eqref{eqn:Abar}
are partitioned accordingly as
\begin{equation}
 A_{p}
    = \begin{bmatrix}
         A_{p,\Ccal\Ccal} & A_{p,\Ccal\Ncal}\\
         A_{p,\Ncal\Ccal} & A_{p,\Ncal\Ncal}
       \end{bmatrix},~~~
 \overline{A} 
   = \begin{bmatrix}
       \overline{A}_{\Ccal\Ccal} & \overline{A}_{\Ccal\Ncal}\\
       \overline{A}_{\Ncal\Ccal} & \overline{A}_{\Ncal\Ncal}
       \end{bmatrix}.
\label{eqn:Abar:part}
\end{equation}

Since the time average $y_{\Ccal}$ has converged sufficiently
by time $k_0$, the proposed approach employs the value $y_{\Ccal}(k_0)$ 
as $x_{\Ccal}(k)$ for all $k\geq k_0$.
Hence, the value at time $k_0$ is reset as
\[
  x(k_0)
  = \begin{bmatrix}
      y_{\Ccal}(k_0)\\
      x_{\Ncal}(k_0)
    \end{bmatrix}.
\]
The updates are carried out through the distributed 
algorithm given by
\begin{equation}
  x(k+1) 
   = \widetilde{A}_{\eta(k)} x(k) + \frac{\hat{m}}{n} \tilde{s},
     ~~~~k\geq k_0, 
 \label{eqn:approx:dist:x}
\end{equation}
where
\begin{align}
 \widetilde{A}_{\eta(k)} 
   &= \begin{bmatrix}
       I & 0\\
       \widetilde{A}_{\eta(k), \Ncal\Ccal} 
            & \widetilde{A}_{\eta(k),\Ncal\Ncal}
     \end{bmatrix}
   := \begin{bmatrix}
       I & 0\\
       (1-\hat{m})A_{\eta(k), \Ncal\Ccal} 
         & (1-\hat{m})$ $A_{\eta(k),\Ncal\Ncal}
     \end{bmatrix},
 \tilde{s}
  := \begin{bmatrix}
       0\\
       \one
     \end{bmatrix},
    \label{eqn:approx:dist:Aeta}
\end{align}
with $\one\in\R^{n-n_1(k_0)}$.
We note that the matrices $\widetilde{A}_{p}$ are
nonnegative, but are no longer stochastic; the sums of the 
entries of the first $n_1(k_0)$ columns are larger than 1 while those
of the other columns are smaller than 1. 
Hence, though $x(k)\geq 0$ still holds,
the state $x(k)$ may not be a probability vector.  
In addition, this scheme is in the distributed form of \eqref{eqn:xAeta}, 
and not the one in \eqref{eqn:xMmod} based on the modified link
matrices.

The time average $y(k)$ is also modified 
by fixing the entries for $i\in\Ccal(k_0)$ as
\[
 y(k) 
   = \begin{bmatrix}
       y_{\Ccal}(k_0)\\ 
       y_{\Ncal}(k)
     \end{bmatrix},~~~~k\geq k_0,
\]
where $y_{\Ncal}(k)$ is determined through the
original formula \eqref{eqn:yk}.

For the approximate update scheme \eqref{eqn:approx:dist:x}, 
its average state $\overline{x}(k):=E[x(k)]$ follows
the recursion
\begin{equation}
  \overline{x}(k+1)
    = \widehat{A}\; \overline{x}(k) + \frac{\hat{m}}{n}\tilde{s},
     ~~~k\geq k_0,
  \label{eqn:approx:xave}
\end{equation}
where the average link 
matrix $\widehat{A}:= E[\widetilde{A}_{\eta(k)}]$ 
is given by
\begin{equation}
 \widehat{A}
    = \begin{bmatrix}
       I & 0\\
       \widehat{A}_{\Ncal\Ccal} 
         & \widehat{A}_{\Ncal\Ncal} 
     \end{bmatrix}
    := \begin{bmatrix}
       I & 0\\
       (1-\hat{m})\overline{A}_{\Ncal\Ccal} 
          & (1-\hat{m})\overline{A}_{\Ncal\Ncal}
     \end{bmatrix}.
\label{eqn:Mtilbar}
\end{equation}

Regarding this average link matrix, 
the following result will be
useful in the subsequent development.

\begin{lemma}\label{lem:Ann}\rm
The submatrix $\widehat{A}_{\Ncal\Ncal}$ of the average
link matrix $\widehat{A}$ as given in \eqref{eqn:Mtilbar} 
satisfies the following:
\begin{enumerate}
\item[(i)]  $\rho(\widehat{A}_{\Ncal\Ncal})\in[0,1-\hat{m}]$ and,
in particular, $\widehat{A}_{\Ncal\Ncal}$ is a stable matrix. 
\item[(ii)] $(I-\widehat{A}_{\Ncal\Ncal})^{-1}\geq 0$.
\end{enumerate}
\end{lemma}

\Proofit
(i)~Since the original average link matrix $\overline{A}$ 
in \eqref{eqn:Abar:part} is a stochastic matrix, 
the block diagonal matrix $\diag(0,\overline{A}_{\Ncal\Ncal})$
containing the submatrix $\overline{A}_{\Ncal\Ncal}$ satisfies
$\overline{A}\geq \diag(0,\overline{A}_{\Ncal\Ncal})\geq 0$.
By the property of nonnegative matrices \cite{HorJoh:85}, it follows that
$1=\rho(\overline{A})\geq \rho(\overline{A}_{\Ncal\Ncal})\geq 0$.
Therefore, $\rho(\widehat{A}_{\Ncal\Ncal})=
\rho((1-\hat{m})\overline{A}_{\Ncal\Ncal})\in[0,1-\hat{m}]$.
By Lemma~\ref{lem:Mbar2}~(i), $\hat{m}\in(0,1)$ and hence
$\rho(\widehat{A}_{\Ncal\Ncal})<1$.

(ii)~Let $\lambda:=1-\rho(\widehat{A}_{\Ncal\Ncal})$.
This is the eigenvalue of $I-\widehat{A}_{\Ncal\Ncal}$ 
with the smallest real part.
By (i), we have $\lambda>0$. 
Thus, $I-\widehat{A}_{\Ncal\Ncal}$ is an $M$-matrix%
\footnote{A matrix $X\in\R^{n\times n}$ is said to be 
an $M$-matrix if its off-diagonal entries are nonpositive 
and all eigenvalues have positive real parts.},
so it has an inverse that is nonnegative \cite{HorJoh:91}.
\mbox{}\EndProof

\smallskip
We remark that in (i) in the lemma, the level of stability is 
affected by the parameter $\alpha$
as it determines the size of $\hat{m}$.

It is clear that 
the value vector $x^*$ (with the coordinate change)
is an equilibrium of the system \eqref{eqn:approx:xave}.
We partition it as
\begin{equation}
 x^* = \begin{bmatrix}
         x_{\Ccal}^*\\
         x_{\Ncal}^*
       \end{bmatrix}.
\label{eqn:approx:xast}
\end{equation}
It is also simple to show that
the vector $\widetilde{x}(k_0)$ given by 
\begin{align}
 \widetilde{x}(k_0)
  &= \begin{bmatrix}
        \widetilde{x}_{\Ccal}(k_0)\\
        \widetilde{x}_{\Ncal}(k_0)
      \end{bmatrix}
  := \begin{bmatrix}
       y_{\Ccal}(k_0)\\
       \bigl(
         I - \widehat{A}_{\Ncal\Ncal}
       \bigr)^{-1} 
         \bigl(
           \widehat{A}_{\Ncal\Ccal}\;
             y_{\Ccal}(k_0) + \frac{\hat{m}}{n}s
         \bigr)
     \end{bmatrix}
\label{eqn:approx:xdash2}
\end{align}
is an equilibrium of the system \eqref{eqn:approx:xave}.
This vector always exists by (ii) in the lemma above. 

After the pages in $\Ccal(k_0)$ have terminated their updates,
the dynamics of the scheme can be characterized as follows.

\begin{lemma}\label{lem:approx:xdash2}\rm
For the distributed update scheme \eqref{eqn:approx:dist:x}
and its average system \eqref{eqn:approx:xave}, 
the following statements hold.
\begin{enumerate}
\item[(i)] The average state $\overline{x}(k)$ converges to 
$\widetilde{x}(k_0)$ and, in particular,
$\overline{x}_{\Ncal}(k)\rightarrow \widetilde{x}_{\Ncal}(k_0)$ 
as $k\rightarrow\infty$.
\item[(ii)]If 
$\abs{\widetilde{x}_{\Ccal}(k_0)-x_{\Ccal}^*}\leq\delta x_{\Ccal}^*$, 
then $\abs{\widetilde{x}_{\Ncal}(k_0)-x_{\Ncal}^*}\leq \delta x_{\Ncal}^*$.
\end{enumerate}
\end{lemma}

\smallskip
\Proofit
(i)~Since $\widetilde{x}(k_0)$ is an equilibrium of 
the average system \eqref{eqn:approx:xave}, it follows that
\begin{align*}
 \overline{x}(k+1) - \widetilde{x}(k_0) 
  &= \begin{bmatrix}
      I & 0\\
      \widehat{A}_{\Ncal\Ccal} & \widehat{A}_{\Ncal\Ncal}
    \end{bmatrix} 
      \bigl(
         \overline{x}(k)-\widetilde{x}(k_0)
      \bigr).
\end{align*}
Here, we have $\overline{x}_{\Ccal}(k) - \widetilde{x}_{\Ccal}(k_0)=0$,
$k\geq k_0$. 
Thus,
\begin{align*}
 \overline{x}_{\Ncal}(k+1) - \widetilde{x}_{\Ncal}(k_0) 
    &= \widehat{A}_{\Ncal\Ncal} 
          \left(
             \overline{x}_{\Ncal}(k) - \widetilde{x}_{\Ncal}(k_0) 
          \right).
\end{align*}
By Lemma~\ref{lem:Ann}\;(i),
$\widehat{A}_{\Ncal\Ncal}$ is a stable matrix. 
Hence, we have
$\overline{x}_{\Ncal}(k) - \widetilde{x}_{\Ncal}(k_0) \rightarrow 0$ 
as $k\rightarrow\infty$.

(ii)~For the average system \eqref{eqn:approx:xave}, 
$\widetilde{x}(k_0)$ and $x^*$ are both equilibria, and 
thus
\[
  \widetilde{x}_{\Ncal}(k_0) - x_{\Ncal}^*
   = \bigl(
        I - \widehat{A}_{\Ncal\Ncal}
     \bigr)^{-1} \widehat{A}_{\Ncal\Ccal} 
           \bigl(
              \widetilde{x}_{\Ccal}(k_0) - x^*_{\Ccal}
           \bigr).
\]
By Lemma~\ref{lem:Ann}\;(ii), 
$(I - \widehat{A}_{\Ncal\Ncal})^{-1}\geq 0$.
Moreover, by construction, $\widehat{A}_{\Ncal\Ccal}\geq 0$.  
Thus, using the assumption, we have
\begin{align*}
 \left|
    \widetilde{x}_{\Ncal}(k_0) - x_{\Ncal}^*
 \right|
  &\leq \bigl(
            I - \widehat{A}_{\Ncal\Ncal}
        \bigr)^{-1} \widehat{A}_{\Ncal\Ccal} \;
           \left| 
             \widetilde{x}_{\Ccal}(k_0) - x^*_{\Ccal}
           \right| \\
  &\leq \bigl(
          I - \widehat{A}_{\Ncal\Ncal}
       \bigr)^{-1} \widehat{A}_{\Ncal\Ccal} \;
           \delta x^*_{\Ccal}\\
  &\leq \delta 
          \bigl(
             I - \widehat{A}_{\Ncal\Ncal}
          \bigr)^{-1} 
          \biggl(
             \widehat{A}_{\Ncal\Ccal} \; x^*_{\Ccal} 
               + \frac{\hat{m}}{n}\one
          \biggr)
  = \delta x^*_{\Ncal},
\end{align*}
where the last equality follows because
$x^*$ is an equilibrium of \eqref{eqn:approx:xave}.
\EndProof

\smallskip
The lemma shows that if the values in 
$y_{\Ccal}(k_0)=\widetilde{x}_{\Ccal}(k_0)$ are 
actually close to the true values $x^*_{\Ccal}$, then 
via the recursion in \eqref{eqn:approx:dist:x},
we can still obtain an approximate value
$\widetilde{x}_{\Ncal}(k_0)$ in the average sense
for all other states; the 
approximation level is the same as that for $\widetilde{x}_{\Ccal}(k_0)$,
represented by the parameter $\delta$.

The following is the main convergence result 
for the scheme described in this section.

\begin{theorem}\label{thm:approx:erg}\rm
Consider the distributed scheme in \eqref{eqn:approx:dist:x},
where under the given sample paths $\{\eta_i(k)\}_{k=0}^{k_0-1}$,
$i=1,\ldots,n$, at time $k_0$, the updates at
$n_1(k_0)$ pages have terminated.
The time average $y_{\Ncal}(k)$, $k\geq k_0$,
converges to the equilibrium $\widetilde{x}_{\Ncal}(k_0)$
in the mean-square sense as 
$E\bigl[
   \bigl\|
      y_{\Ncal}(k) - \widetilde{x}_{\Ncal}(k_0)
   \bigr\|^2
  \bigr] \rightarrow 0$, $k\rightarrow\infty$.
\end{theorem}

The proof is presented in Appendix~\ref{sec:app:C}. 
It is based on that for Theorem~\ref{thm:erg}. 
However, unlike the setup there, the distributed
link matrices in the current scheme are not stochastic. 
This means that we cannot employ general Markov process results of, 
e.g., \cite{Cogburn:86}. 
In contrast, the proof relies 
on the stability of the submatrix $\widehat{A}_{\Ncal\Ncal}$ 
as shown in Lemma~\ref{lem:Ann}\;(i).
Also, for this reason, the analysis does not involve
the modified link matrices
such as $M_{\eta_1(k),\ldots,\eta_n(k)}$ 
that appeared in the previous section. 

\subsection{Distributed algorithm with update termination}

We present the distributed algorithm 
with update termination based on the results from this section. 

\begin{algorithm}\rm\label{alg:dist}
For $i\in\Vcal$, page $i$ executes the following.
\begin{enumerate}
\item[0)] Initialize the parameters
$n$, $\alpha$, $x_i(0)$, $N_s$, and $\delta$.
Set $k=0$, $\Ccal(0)=\emptyset$, and $n_1(0)=0$.

\item[1)] At time $k$, generate $\eta_i(k)\in\{0,1\}$ under 
the probability $\alpha$.
If $\eta_i(k)=1$, then send the value $x_i(k)$ to
pages $j\nin\Ccal(k)$ that it is linked to, and
request pages $j\nin\Ccal(k)$ having links to 
it for their values.

\item[2)] Update the value $x_i(k)$ and its time average by
\begin{equation}
\begin{split}
  x_i(k+1) 
  &= \sum_{j=1}^n
        \bigl(
           \widetilde{A}_{\eta(k)} 
        \bigr)_{ij}\; x_j(k)
             + \frac{\hat{m}}{n},\\
 y_i(k) 
  &= \frac{1}{k+1} \sum_{\ell=0}^{k}x_i(\ell),
\end{split}
\label{eqn:alg:dist:x}
\end{equation}
where 
$\widetilde{A}_{\eta(k)}$
is constructed by \eqref{eqn:approx:dist:Aeta} using $\Ccal(k)$.

\item[3)] Check if $y_i(k)$ has sufficiently converged based on 
\eqref{eqn:yconv}.  If so, then 
(i) add $i$ to the set $\Ccal(k)$,
(ii) send $y_i(k)$ to the pages having direct links
to page $i$, and (iii) 
fix $x_i(\ell)=y_i(\ell)=y_i(k)$ for $\ell\geq k$.

\item[4)] If $\Ccal(k)=\Vcal$, then terminate the algorithm.
Otherwise, set $\Ccal(k+1)=\Ccal(k)$ and $k=k+1$, and then 
go to Step~1.
\End
\end{enumerate}
\end{algorithm}

\vspace*{2mm}
We remark that, in this scheme, 
the choice of the parameters $\delta$ and $N_s$ 
affects the accuracy in the values when the pages terminate 
their updates as well as the time when the pages decide to do so.
Taking $\delta$ smaller and/or $N_s$ larger will improve the
value estimates, but will require longer time before the
updates terminate; this in turn will keep the computation 
and communication load higher. 


\section{Discussion on asynchronous iteration methods}
\label{sec:asynch}

In this section, 
we discuss the application of a numerical 
analysis method known as asynchronous iteration \cite{BerTsi:89} to the 
distributed computation of PageRank values. 
Deterministic algorithms for the PageRank problem
under this approach have been discussed in, e.g., 
\cite{deJBra:07,KolGalSzy:06}.
We present a randomization-based algorithm and
clarify its relation to the schemes of this paper.

Consider the original update scheme in \eqref{eqn:xAeta}
based on the power method. 
Let $\eta_i(k)\in\{0,1\}$ for $i\in\Vcal$, $k\in\Z_+$, 
be the i.i.d.\ random processes whose distributions
are as in \eqref{eqn:alpha}.
Similarly to the scheme with simultaneous updates 
in Section~\ref{sec:simul},
when $\eta_i(k)=1$ at time $k$, then page $i$ initiates 
an update; such an event occurs with probability $\alpha$.
However, the difference is that 
this update is performed as in the power method, and
moreover pages whose corresponding processes $\eta_i(k)$
are zero do not make any updates. 

The distributed update recursion is given as follows:
\begin{equation}
  \check{x}(k+1) = \check{M}_{\eta_1(k),\ldots,\eta_n(k)} 
                        \check{x}(k),
\label{eqn:xasynch}
\end{equation}
where the initial state $\check{x}(0)$ is a probability vector
and the link matrices are given by
\begin{align}
 \bigl(
     \check{M}_{p_1,\ldots,p_n}
 \bigr)_{ij}
  := \begin{cases}
       (1-m)a_{ij}+ \frac{m}{n} & \text{if $p_i = 1$},\\
       1                       & \text{if $p_i = 0$ and $i=j$},\\
       0                       & \text{otherwise}.
     \end{cases}
\end{align}
It is clear that these matrices keep the rows
of the original link matrix $M$ in \eqref{eqn:M} 
for the pages that initiate updates.
Other pages just keep their previous values. 
Thus, these matrices are not stochastic. 

The following result shows that through this algorithm,
we can compute the PageRank values.  

\begin{lemma}\label{lem:asynch}\rm
Under the distributed update scheme of \eqref{eqn:xasynch},
for every initial state $x(0)$ that is a probability vector,
the PageRank value $x^*$ is obtained as
$\check{x}(k)\rightarrow x^*$, $k\rightarrow\infty$, with 
probability one. 
\end{lemma}

The distributed update scheme in \eqref{eqn:xasynch} 
is a randomized version of the one 
in \cite[Section~6.2]{BerTsi:89}, and the proof can be
extended in a straightforward way. 
Specifically, it relies on the property $\rho(M)=1$ 
that the link matrix $M$ has. 
The algorithm is based on general asynchronous iteration 
algorithms for distributed computation of fixed points
in the field of numerical analysis.
It is interesting to note that the proof of the result above 
employs an argument similar to that of Lyapunov functions. 
We also point out that the convergence rate is exponential and
the general scheme can handle delays in the communication. 

In comparison, the algorithms proposed in this paper 
have the following characteristic features. 
First, the link matrices in the update schemes \eqref{eqn:xMi1}
and \eqref{eqn:xAeta} are stochastic, 
and this property is exploited in the convergence analysis. 
It moreover provides the relation to consensus type problems
as discussed in Section~\ref{sec:consensus}. 
Second, there is a difference with regard to the type of
links over which communication takes place.
In particular, it is shown in a subsequent paper 
\cite{IshTemBaiDab:09} 
that the present approach can be extended 
in such a way that each page
communicates only with those connected by outgoing links;
the information of such links are by default available locally. 
By contrast, in the asynchronous iteration algorithm,
pages must utilize the incoming links. This means that
popular pages linked from many pages need extra 
storage to keep the data of such links.



\section{Numerical example}
\label{sec:example}

We present an example with 1,000 web pages ($n=1,000$).
The links among the pages were randomly generated. 
The first ten pages are designed to have high rankings
and are given links from over 90\% of the pages. 
Others have between 2 and 333 links per page. 

\begin{figure}[t]
  \centering
  \fig{8cm}{5cm}{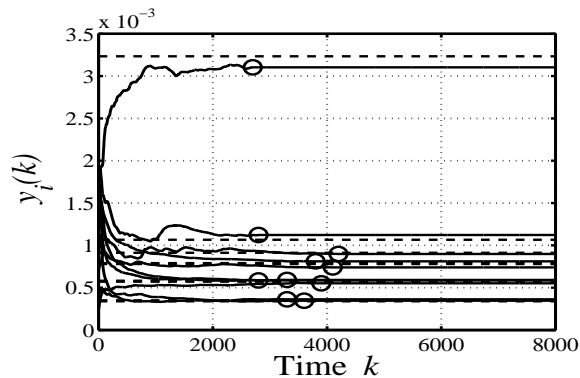}
  \vspace*{-2mm}
  \caption{Sample paths of $y_i$ (solid lines) with
           the times at which updates stopped 
           (marked by $\bigcirc$ for each page)
           and the true PageRank values $x^*_i$ (dashed lines)
           for $i=21,\ldots,30$.}
  \label{fig:y_term}
\end{figure}

\begin{figure}[t]
  \centering
  \fig{8cm}{5cm}{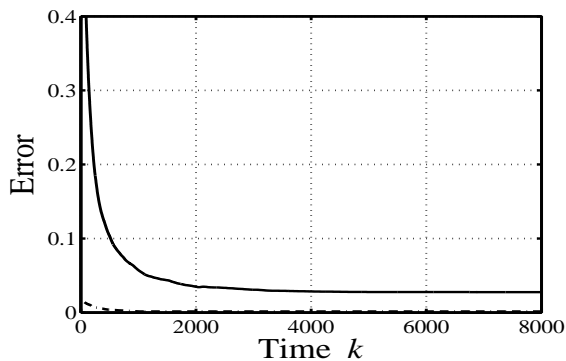}
  \vspace*{-2mm}
  \caption{Estimation error:
           $\norm{e(k)}_{1}$ (solid line) and
           $\norm{e(k)}_{\infty}$ (dash-dot line)}
  \label{fig:err_term}
\end{figure}

We ran Algorithm~\ref{alg:dist}
in Section~\ref{sec:approx} where each page 
initiates an update at a fixed probability $\alpha=0.01$ and 
terminates the updates when an approximate value
is obtained.  
In the distributed scheme \eqref{eqn:alg:dist:x}, 
we generated sample paths of the processes $\eta_i(k)$, 
$i\in\Vcal$, which determine the pages initiating updates,
and computed the state $x(k)$.
The initial state $x(0)$ was taken as a random probability vector.
The parameters for the update termination were chosen as follows:
The number $N_s$ of steps before stopping the update was $N_s=800$ and 
the parameter $\delta$ determining the level of approximation was
$\delta=0.01$.  
We chose these values so that the characteristics
of this scheme are visible in the plots. 

The responses of the time average $y_i$ for $i=21,\ldots,30$ 
are shown in Fig.~\ref{fig:y_term}. 
The time when the corresponding pages terminated their updates
are marked by $\bigcirc$.  
We observe that the convergence is fairly fast, and 
the updates stop by time $k=4,500$ for these pages. 

Let the errors in the estimates be $e(k):=y(k)-x^*$.
In Fig.~\ref{fig:err_term}, these are shown for two cases:
$\norm{e(k)}_{1}$ under the $\ell_1$ norm and
$\norm{e(k)}_{\infty}$ under the $\ell_{\infty}$ norm.
The plot shows that the error in the individual values of
$y_i$ (measured by the $\ell_{\infty}$ norm)
rapidly decreases and remains small
while the total error (in the $\ell_1$ norm)
also decreases but at a slower rate.

\begin{figure}[t]
  \centering
  \fig{8cm}{5cm}{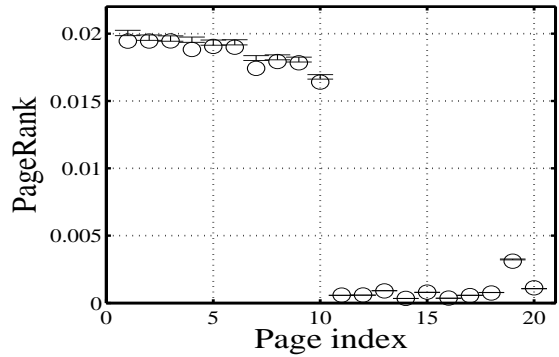}
  \vspace*{-2mm}
  \caption{Ranges of approximate PageRank values 
          (marked by two lines) and 
     $y_i$ for $i=1,\ldots,20$, at $k=8,000$ (marked as $\bigcirc$)}
  \label{fig:pr_term}
\end{figure}

In Fig.~\ref{fig:pr_term}, the final values of $y_i$ for the
first twenty pages are plotted as $\bigcirc$ together 
with the acceptable ranges of error, that is,
$[(1-\delta)x^*_i,(1+\delta)x^*_i]$ by two lines connected in the middle.
As we mentioned in Section~\ref{sec:approx}, the time average 
$y$ is no longer normalized in this case.  However, 
the sum of all $y_i$ at $k=8,000$ turned out to be about $0.989$, which is 
very close to the desired value 1.

\section{Conclusion}
\label{sec:concl}

In this paper, we first gave an overview of
the PageRank computation problem, which is critical
in making accurate search results with Google.
We introduced a randomization-based distributed approach 
for the computation of PageRank values and 
showed the mean-square convergence of the proposed schemes.
It was demonstrated that the approach has a clear
relation to consensus type problems.
The algorithms were generalized 
in the recent papers \cite{IshTem_acc:09,IshTemBaiDab:09},  
where random link failures and computations based on aggregating
groups of pages are addressed and more discussions on the 
advantages of this approach can be found.
Future research will deal with the effects of communication
delays and the improvement of convergence rate, 
and also study issues related to implementation
of the proposed distributed algorithms.

\smallskip
\noindent
{\it Acknowledgement}:~We are thankful to 
Er-Wei Bai, B.~Ross Barmish, Tamer Ba\c{s}ar, 
Fabrizio Dabbene, Soura Dasgupta, Shinji Hara, 
Zhihua Qu, M.~Vidyasagar, and Yutaka Yamamoto
for their helpful comments and discussions
on this work.  



{\small

}


\appendix

\section{Appendix}

In this appendix, following some preliminary material,
the proofs of Theorems~\ref{thm:erg} and
\ref{thm:approx:erg} are given.

\subsection{Preliminaries}

We present some results related to infinite 
products of stochastic matrices from \cite{Seneta:81,Hartfiel:98}.
These are required for the proof of 
Theorem~\ref{thm:erg} given in the next subsection. 

First, the notion of weak ergodicity is introduced.

\begin{definition}\rm
Given a sequence of stochastic matrices 
$\{P(k)\}_{k=0}^{\infty}\subset\R^{n\times n}$,
let their (backward) product be $T(k):=P(k)\cdots P(0)$.  
The sequence $\{P(k)\}_{k=0}^{\infty}$ is said to be
weakly ergodic if
\begin{equation}
   t_{ri}(k) - t_{rj}(k) \rightarrow 0,~~
          k\rightarrow \infty,~~\forall i,j,r\in\{1,\ldots,n\}.
  \label{eqn:ergodic}
\end{equation}
\end{definition}

\vspace*{3mm}
In \eqref{eqn:ergodic}, $t_{ri}(k)$ for all $i$ tend to 
be equal as $k\rightarrow\infty$, that is,
all columns of the product matrix $T(k)$ coincide in the limit.
However, in general, the columns 
do not converge to a single vector.  

To characterize matrix sequences that are weakly ergodic,
we employ the tool known as the coefficient of ergodicity.  
Let $\tau(\cdot)$ be the scalar function
for stochastic matrices in $\R^{n\times n}$ given by
\begin{equation}
  \tau(P) 
   := \frac{1}{2}\max_{i,j}\sum_{r=1}^{n}\abs{p_{ri}-p_{rj}}.
\label{eqn:tau}
\end{equation}

This function $\tau(\cdot)$ has the following properties.

\begin{lemma}\label{lem:coef}\rm
\begin{enumerate}
\item[(i)] $\tau(P)\in[0,1]$ and, moreover, $\tau(P)=0$ 
if and only if there exists a probability vector 
$v\in\R^n$ such that $P=v \one^T$, where $\one=[1\;\cdots\;1]^T\in\R^n$.

\item[(ii)]
$\tau(P) 
= \max\{
       \norm{Px}_1:x\in\R^n,\norm{x}_1=1,\sum_{i}x_i=0
      \}$.

\item[(iii)] $\tau(PQ)\leq \tau(P)\tau(Q)$ for stochastic matrices $P,Q$.
\end{enumerate}
\end{lemma}




\vspace*{2mm}
Weak ergodicity can be
characterized by the following lemma.

\begin{lemma}\label{lem:weakerg}\rm
 For a sequence of stochastic matrices $\{P(k)\}_{k=0}^{\infty}$,
 their product $P(k)\cdots P(0)$ is weakly ergodic
 if $\tau(P(k))\leq \tau_0$ for all $k$, where $\tau_0\in(0,1)$ is a scalar.
\end{lemma}

\subsection{The distributed update scheme}
\label{sec:app:B}

We now analyze the proposed algorithm in Section~\ref{sec:dist1}.

\begin{lemma}\label{lem:tauM}\rm
For the distributed update scheme \eqref{eqn:xM} and 
its average system \eqref{eqn:xMbar}, the following hold.
\begin{enumerate}
\item[(i)] 
The matrices $M_i$, 
$i\in\{1,\ldots,n\}$, and $\overline{M}$ satisfy 
$\tau(M_i) < 1-\hat{m}$ and $\tau(\overline{M}) < 1-\hat{m}$.
\item[(ii)] For any mode sequence $\{\theta(k)\}$,
the matrix sequence $\{M_{\theta(k)}\}$ is weakly ergodic.
\end{enumerate}
\end{lemma}

\vspace*{2mm}
\Proofit
(i)~~We show only for $\overline{M}$ since the case of $M_i$ is similar.
Recall that by \eqref{eqn:Abar0} and \eqref{eqn:Mi}, 
the average matrix $\overline{M}$ can be expressed as
$\overline{M}= (1-\hat{m}) \overline{A} + \frac{\hat{m}}{n} S$.
Thus, we have
\begin{align*}
  \tau(\overline{M}) 
   &= \frac{1}{2}\max_{i,j}\sum_{r=1}^{n}
         \left|
             (\overline{M})_{ri} - (\overline{M})_{rj}
         \right|\\
   &= (1-\hat{m})\frac{1}{2}\max_{i,j}\sum_{r=1}^{n}
         \left|
             (\overline{A})_{ri} - (\overline{A})_{rj}
         \right|
   = (1-\hat{m})\tau(\overline{A}).
\end{align*}
The matrix $\overline{A}$ is stochastic, and hence,
by Lemma~\ref{lem:coef}~(i), it holds that $\tau(\overline{A})\leq 1$.
Consequently, we arrive at the inequality $\tau(\overline{M})\leq 1-\hat{m}$.

(ii)~~This follows from (i) and Lemma \ref{lem:weakerg} since
$\hat{m}\in(0,1)$ by Lemma~\ref{lem:Mbar}~(i). 
\EndProof

\smallskip
{\it Proof of Theorem~\ref{thm:erg}:}~~
Let the error from the average be
$e(k) := x(k) - x^*$.  
Note that $e(k)$ satisfies $\sum_{i=1}^n e_i(k)=0$.
This is because in the systems \eqref{eqn:xM} and \eqref{eqn:xMbar},
by assumption, the initial states
are probability vectors,
and furthermore, $M_i$, $i=1,\ldots,n$, and $\overline{M}$ are 
stochastic matrices; hence, both $x(k)$ and $x^*$
are nonnegative vectors whose entries add up to 1.

Observe that
\begin{align*}
 y(k) - x^*
  &= \frac{1}{k+1}\sum_{\ell=0}^{k} (x(\ell) - x^*)
   = \frac{1}{k+1}\sum_{\ell=0}^{k} e(\ell).
\end{align*}
Thus, 
\begin{align}
 E[\norm{y(k) - x^*}^2]
  &= E\biggl[
        \biggl\|
          \frac{1}{k+1}\sum_{\ell=0}^{k} e(\ell)
        \biggr\|^2
       \biggr] 
  = \frac{1}{(k+1)^2}
      E\biggl[
        \sum_{\ell=0}^{k} e(\ell)^T e(\ell) 
          + 2 \sum_{\ell=0}^{k-1} \sum_{r=1}^{k-\ell} e(\ell)^T e(\ell+r)
      \biggr]\notag\\
  &= \frac{1}{(k+1)^2}
       \biggl\{
        \sum_{\ell=0}^{k} E[e(\ell)^T e(\ell)] 
          + 2 \sum_{\ell=0}^{k-1} \sum_{r=1}^{k-\ell} E[e(\ell)^T e(\ell+r)]
      \biggr\}.
\label{eqn:thm:erg:2}
\end{align}
We use the norm relation $\norm{z}\leq\norm{z}_1$ 
for $z\in\R^n$ \cite{HorJoh:85} and the property
$\norm{x(\ell)}_1=\norm{x^*}_1=1$ 
to obtain the bound $\norm{e(\ell)}\leq 2$.
Then, in the first summation term in \eqref{eqn:thm:erg:2}, 
we have
\begin{equation}
  \sum_{\ell=0}^{k} E[e(\ell)^T e(\ell)] 
     \leq 4(k+1).
\label{eqn:thm:erg:2a}
\end{equation}
In the second summation term, 
we see that the summands can be written as
\begin{align}
 E[e(\ell)^T e(\ell+r)] 
 &= E[ e(\ell)^T (x(\ell+r) - x^*)] \notag\\
 &= E\bigl[
       e(\ell)^T 
        \bigl(
           M_{\theta(\ell+r-1)}\cdots M_{\theta(\ell)} x(\ell) - x^*
        \bigr)
       \bigr].
 \label{eqn:thm:erg:3}
\end{align}
Here, by taking the expectation of the matrix 
product $M_{\theta(\ell+r-1)}\cdots M_{\theta(\ell)}$ with respect to
the random variables $\theta(\ell+r-1),\ldots,\theta(\ell)$, 
\begin{align}
E[e(\ell)^T e(\ell+r)] 
 &= E\bigl[
       e(\ell)^T 
        \bigl(
          E[M_{\theta(\ell+r-1)}\cdots M_{\theta(\ell)}] x(\ell) - x^*
        \bigr)
       \bigr] \notag\\
 &= E\bigl[
       e(\ell)^T 
        \bigl(
          E[M_{\theta(\ell+r-1)}]\cdots E[M_{\theta(\ell)}] x(\ell)  - x^*
        \bigr)
       \bigr] \notag\\
 &= E\bigl[
       e(\ell)^T 
          ( \overline{M}^r x(\ell) - x^* )
       \bigr],
 \label{eqn:thm:erg:3b}
\end{align}
where the second and third equalities follow
from the independence of $\theta(\ell+r-1),\ldots,\theta(\ell)$ and
the definition of the average matrix $\overline{M}$,  
respectively.
Since, by Lemma~\ref{lem:Mbar}~(iii), 
$x^*$ is the eigenvector of $\overline{M}$ 
for the eigenvalue 1, it follows that 
$\overline{M}^r x(\ell) - x^*=\overline{M}^r (x(\ell) - x^*)$. 
Further, we have $x(\ell)-x^*=e(\ell)$ and 
again apply the fact $\norm{z}\leq\norm{z}_1$, $z\in\R^n$,
to derive from \eqref{eqn:thm:erg:3b} that
\begin{align}
 E[e(\ell)^T e(\ell+r)] 
 &= E\bigl[
        e(\ell)^T \overline{M}^r e(\ell)
     \bigr]
 \leq E\bigl[
           \norm{e(\ell)} \norm{\overline{M}^r e(\ell)}
        \bigr] \notag\\
 &\leq  2\;
        E\bigl[
           \norm{\overline{M}^r e(\ell)}_1
        \bigr],
 \label{eqn:thm:erg:3a}
\end{align}
where in the last inequality, 
we also used $\norm{e(\ell)}\leq 2$.
As we have mentioned above, it holds that $\sum_{i=1}^n e_i(\ell)=0$.
Thus, apply Lemma~\ref{lem:coef} (ii) and (iii) 
to \eqref{eqn:thm:erg:3a} and obtain
\begin{align}
 E[e(\ell)^T e(\ell+r)] 
 &\leq  2\;
        \tau(\overline{M}^r)
       E\bigl[
          \norm{e(\ell)}_1
       \bigr]\notag\\
 &\leq 2\;\tau(\overline{M})^r 
        E\bigl[
           \norm{e(\ell)}_1
         \bigr]
 \leq 4\;\tau(\overline{M})^r,
 \label{eqn:thm:erg:6}
\end{align}
where the last inequality is due to $\norm{e(\ell)}_1\leq 2$.
Note that by Lemma~\ref{lem:tauM}, $\tau(\overline{M})<1$.

Finally, by substituting \eqref{eqn:thm:erg:2a} and
\eqref{eqn:thm:erg:6} into \eqref{eqn:thm:erg:2}, we have
\begin{align*}
 E[\norm{y(k) - x^*}^2]
  &\leq \frac{1}{(k+1)^2}
       \biggl\{
         4(k+1)
          + 2\sum_{\ell=0}^{k-1} 
                \sum_{r=0}^{k-\ell} 
                  4\, \tau(\overline{M})^r
      \biggr\}\notag\\
  &\leq \frac{4}{k+1}
         \biggl( 
           1 + \frac{2}{1-\tau(\overline{M})} 
         \biggr), 
\end{align*}
and hence using the bound on $\tau(\overline{M})$ in Lemma~\ref{lem:tauM}~(i),
we obtain
\begin{align}
 E[\norm{y(k) - x^*}^2]
  &\leq \frac{4(2 + \hat{m})}{\hat{m}(k+1)}
  \rightarrow 0,~~~k\rightarrow\infty.
\label{eqn:thm:erg:conv}
\end{align}
Thus, 
the PageRank value $x^*$ is obtained through the time average $y$.
\EndProof

\subsection{Proof of Theorem~\ref{thm:approx:erg}}
\label{sec:app:C}

For simplicity, let the initial time of the
update scheme \eqref{eqn:approx:dist:x} to be $k_0+1 =0$. 
Further, we write $\tilde{x}_{\Ncal}$ for $\tilde{x}_{\Ncal}(k_0)$.
Denote the error between the state and the average by
$e(k) := x_{\Ncal}(k) - \widetilde{x}_{\Ncal}$.  Then, 
\begin{align*}
 y_{\Ncal}(k) - \widetilde{x}_{\Ncal}
  &= \frac{1}{k+1}\sum_{\ell=0}^{k} 
            (x_{\Ncal}(\ell) - \widetilde{x}_{\Ncal})
   = \frac{1}{k+1}\sum_{\ell=0}^{k} e(\ell).
\end{align*}
Thus, 
\begin{align}
 E[\norm{y_{\Ncal}(k) - \widetilde{x}_{\Ncal}}^2]
  &= \frac{1}{(k+1)^2}
       \biggl\{
        \sum_{\ell=0}^{k} E[e(\ell)^T e(\ell)] 
         + 2 \sum_{\ell=0}^{k-1} \sum_{r=1}^{k-l} E[e(\ell)^T e(\ell+r)]
      \biggr\}.
\label{eqn:prop:approx:erg:2}
\end{align}
In what follows, we must evaluate the two summation terms on the
right-hand side.

First, we claim that $e(k)$ is uniformly bounded and in particular,
for each $k\geq 0$,
\begin{equation}
  \norm{e(k)}_1
  \leq \norm{x_{\Ncal}(0)}_1 + \frac{\varepsilon}{\hat{m}} 
           + \norm{\widetilde{x}_{\Ncal}}_1 
   =: \gamma,
  \label{eqn:gamma}
\end{equation}
where 
$\varepsilon := \max_{p\in\{0,1\}^n} \norm{
                    \widetilde{A}_{p, \Ncal\Ccal} \widetilde{x}_{\Ccal}
                      + \hat{m}/n\,\one
                  }_1$.
Notice that
\begin{equation}
  \norm{e(k)}_1
   =  \norm{x_{\Ncal}(k) - \widetilde{x}_{\Ncal}}_1
   \leq \norm{x_{\Ncal}(k)}_1 + \norm{\widetilde{x}_{\Ncal}}_1.
\label{eqn:normek}
\end{equation}
From the distributed update law in \eqref{eqn:approx:dist:x}, 
it easily follows that
\[
  x_{\Ncal}(k+1) 
   = \widetilde{A}_{\eta(k),\Ncal\Ncal} x_{\Ncal}(k) 
      + \widetilde{A}_{\eta(k),\Ncal\Ccal} \widetilde{x}_{\Ccal}
       + \frac{\hat{m}}{n} \one.
\]
By definition,
$\widetilde{A}_{\eta(k),\Ncal\Ncal} = (1-\hat{m})A_{\eta(k),\Ncal\Ncal}$ 
and $A_{\eta(k),\Ncal\Ncal}$ is a submatrix of the stochastic
matrix $A_{\eta(k)}$.  Consequently, we have
$\norm{\widetilde{A}_{\eta(k),\Ncal\Ncal}}_1
   \leq 1 - \hat{m}$.
Using this bound, we obtain
\begin{align*}
 \norm{x_{\Ncal}(k+1)}_1 
 &\leq \bigl\|
          \widetilde{A}_{\eta(k),\Ncal\Ncal}
        \bigr\|_1\cdot 
          \norm{x_{\Ncal}(k)}_1
        + \biggl\|
            \widetilde{A}_{\eta(k),\Ncal\Ccal}\widetilde{x}_{\Ccal}
              + \frac{\hat{m}}{n}\one
          \biggr\|_1\\
 &\leq (1 - \hat{m})\norm{x_{\Ncal}(k)}_1 + \varepsilon.
\end{align*}
Thus, 
\begin{align*}
  \norm{x_{\Ncal}(k)}_1
   &\leq (1 - \hat{m})^{k} \norm{x_{\Ncal}(0)}_1
          + \varepsilon \sum_{\ell=0}^{k-1}(1 - \hat{m})^{\ell}\\
  &\leq \norm{x_{\Ncal}(0)}_1 + \frac{\varepsilon}{\hat{m}}.
\end{align*}
Therefore, substituting this into \eqref{eqn:normek}, 
we have shown \eqref{eqn:gamma}.

Now, with the bound \eqref{eqn:gamma},
the first summation term in \eqref{eqn:prop:approx:erg:2} can be 
upper bounded as
\begin{equation}
  \sum_{\ell=0}^{k} E[e(\ell)^T e(\ell)] 
   \leq (k+1)\gamma^2.
\label{eqn:prop:approx:erg:2a}
\end{equation}

We next look at the the second summation term of \eqref{eqn:prop:approx:erg:2}.
The summands can be written as
\begin{align*}
 &E[e(\ell)^T e(\ell+r)] 
 = E[ e(\ell)^T [0~I] (x(\ell+r) - \widetilde{x})] \notag\\
 &~~= E\biggl[
      e(\ell)^T [0~I]
       \biggl(
        \widetilde{A}_{\eta(\ell+r-1)}\cdots 
                       \widetilde{A}_{\eta(\ell)} x(\ell) 
         + \sum_{j=\ell}^{\ell+r-1}
            \widetilde{A}_{\eta(\ell+r-1)}\cdots \widetilde{A}_{\eta(j+1)} 
            \frac{\hat{m}}{n}\tilde{s}
             - \widetilde{x}
        \biggr)
       \biggr].
\end{align*}
Here, by taking the expectation of the matrix products
$\widetilde{A}_{\eta(\ell+r-1)}\cdots \widetilde{A}_{\eta(\ell+j)}$,
$j=0,1,\ldots,r-1$, with 
respect to the random variables $\eta(\ell+r-1),\ldots,\eta(\ell+j)$,
we have
\begin{align*}
 E[e(\ell)^T e(\ell+r)]
 &= E\biggl[
       e(\ell)^T [0~I]
        \biggl(\hspace*{-0.4mm}
          E\bigl[
              \widetilde{A}_{\eta(\ell+r-1)}\cdots \widetilde{A}_{\eta(\ell)}
           \bigr] x(l) \\
 &\hspace*{2.5cm}\mbox{}
         + \hspace*{-0.5mm}
           \sum_{j=\ell}^{\ell+r-1}
            \hspace*{-0.5mm}
            E\bigl[
                \widetilde{A}_{\eta(\ell+r-1)}
                        \cdots \widetilde{A}_{\eta(j+1)}
             \bigr]
            \frac{\hat{m}}{n}\tilde{s}
              - \widetilde{x}        
        \biggr)\hspace*{-0.3mm}
       \biggr] \notag\\
 &= E\biggl[
       e(\ell)^T [0~I]
        \biggl(\hspace*{-0.4mm}
         E\bigl[
             \widetilde{A}_{\eta(\ell+r-1)}]\cdots 
                          E[\widetilde{A}_{\eta(\ell)}
          \bigr] x(\ell)  \\
 &\hspace*{2.5cm}\mbox{}
         + \hspace*{-0.4mm}
           \sum_{j=\ell}^{\ell+r-1}
            \hspace*{-0.4mm}
            E\bigl[
                \widetilde{A}_{\eta(\ell+r-1)}]
                    \cdots E[\widetilde{A}_{\eta(j+1)}
             \bigr]
            \frac{\hat{m}}{n}\tilde{s}
              - \widetilde{x}
        \biggr)\hspace*{-0.3mm}
       \biggr]\\
 &= E\biggl[
       e(\ell)^T [0~I]
          \biggl( 
            \widehat{A}^r x(\ell) 
              + \hspace*{-0.4mm}
                \sum_{j=\ell}^{\ell+r-1}
                \hspace*{-0.4mm}
                \widehat{A}^{\ell+r-1-j}\frac{\hat{m}}{n}\tilde{s}
              - \widetilde{x} 
          \biggr)\hspace*{-0.3mm}
       \biggr],
\end{align*}
where the second and third equalities, respectively, follow from 
the independence of $\eta(\ell+r-1),\ldots,\eta(\ell+j)$ and 
the definition of $\widehat{A}$ in \eqref{eqn:Mtilbar}.
As $\widetilde{x}$ is an equilibrium of the average system in
\eqref{eqn:approx:xave}, it can be shown that
$E[e(\ell)^T e(\ell+r)] 
 = E\bigl[
       e(\ell)^T [0~I]
         \widehat{A}^r (x(\ell) - \widetilde{x})
       \bigr]$.
Here, we have $x(\ell) - \widetilde{x}=[0^T~e(\ell)^T]^T$ and 
use the norm relation $\norm{z}\leq \norm{z}_1$ for all $z\in\R^n$
\cite{HorJoh:85} to derive
\begin{align}
 E[e(\ell)^T e(\ell+r)] 
 &= E\biggl[
        e(\ell)^T \begin{bmatrix}0 & I\end{bmatrix} \widehat{A}^r 
         \begin{bmatrix}
            0\\  e(\ell)
         \end{bmatrix}
     \biggr]\notag\\
 &= E\bigl[
        e(\ell)^T \widehat{A}_{\Ncal\Ncal}^r e(\ell)
     \bigr] 
 \leq E\bigl[
        \norm{e(\ell)}\cdot
           \bigl\|
             \widehat{A}_{\Ncal\Ncal}^r e(\ell)
           \bigr\|
     \bigr]\notag\\
 &\leq E\bigl[
        \norm{e(\ell)}_1 \cdot
           \bigl\|
             \widehat{A}_{\Ncal\Ncal}^r e(\ell)
           \bigr\|_1
     \bigr] 
 \leq E\bigl[
           \norm{e(\ell)}_1^2\cdot
              \bigl\|
                 \widehat{A}_{\Ncal\Ncal}^r
              \bigr\|_1
        \bigr] \notag\\
 &\leq \gamma^2 \bigl\|
                  \widehat{A}_{\Ncal\Ncal}
                \bigr\|^r_1,
 \label{eqn:prop:approx:erg:3a}
\end{align}
where in the last inequality, we used $\norm{e(\ell)}_1\leq \gamma$
from \eqref{eqn:gamma}.
The average matrix $\widehat{A}_{\Ncal\Ncal}$ can be
bounded as
$\bigl\|\widehat{A}_{\Ncal\Ncal}\bigr\|_1 \leq 1 - \hat{m}$
because $\widehat{A}_{\Ncal\Ncal}=(1-\hat{m})\overline{M}_{\Ncal\Ncal}$,
and $\overline{M}_{\Ncal\Ncal}$ is a submatrix of a stochastic matrix.
Thus, we arrive at
\begin{align}
 E[e(\ell)^T e(\ell+r)] 
 &\leq \gamma^2 \left(
                   1 - \hat{m}
                \right)^r.
 \label{eqn:prop:approx:erg:6}
\end{align}

Finally, by substituting \eqref{eqn:prop:approx:erg:2a} and
\eqref{eqn:prop:approx:erg:6} into \eqref{eqn:prop:approx:erg:2}
and by $\hat{m}\in(0,1)$ from Lemma~\ref{lem:Mbar2}~(i),
\begin{align*}
 &E[\norm{y_{\Ncal}(k) - \widetilde{x}_{\Ncal}}^2]\notag\\
  &~~\leq \frac{1}{(k+1)^2}
       \biggl\{
         (k+1)\gamma^2
          + 2\sum_{\ell=0}^{k-1} \sum_{r=1}^{k-\ell} 
              \gamma^2 \left(
                          1 - \hat{m}
                       \right)^r
      \biggr\} \notag\\
  &~~\leq \frac{\gamma^2}{k+1}
         \biggl( 
           1 + 2 \frac{1-\hat{m}}{\hat{m}} 
         \biggr) 
\end{align*}
The far right-hand side converges to zero as $k\rightarrow \infty$,
which completes the proof.
\EndProof



\end{document}